\documentclass[twocolumn,apj]{aastex62}

\received{22 May 2019}
\accepted{8 September 2019}

\usepackage{natbib}
\usepackage{graphicx}
\usepackage{amsmath,amssymb}
\usepackage{txfonts}
\usepackage{epstopdf}
\usepackage{natbib}
\usepackage{amsmath,amssymb}

\turnoffedits
\newcommand{\be}{\begin{equation}}
\newcommand{\ee}{\end{equation}}
\newcommand{\bdm}{\begin{displaymath}}
\newcommand{\edm}{\end{displaymath}}
\newcommand{\bea}{\begin{eqnarray}}
\newcommand{\eea}{\end{eqnarray}}
\newcommand{\ba}{\begin{align}}
\newcommand{\ea}{\end{align}}

\newcommand{\PD}[2]{\partial_{#2} #1}
\newcommand{\TD}[2]{\frac{\mathrm{d} #1}{\mathrm{d}#2} }
\renewcommand{\vec}[1]{{\mbox{\boldmath${\mathrm{#1}}$} }}  %vector
\newcommand{\DIV}[1]{\nabla\cdot {#1}}
\newcommand{\CURL}[1]{\nabla \times {#1}}

\newcommand{\rP}{\mathrm{P}}
\newcommand{\rR}{\mathrm{R}}
\newcommand{\rB}{\mathrm{B}}
\newcommand{\rI}{\mathrm{I}}

\newcommand{\bB}{\vec{B}} 
 
\renewcommand{\bv}{\vec{v}} 
\newcommand{\bJ}{\vec{J}} 
\newcommand{\bey}{\vec{\hat{y}}}
\newcommand{\bez}{\vec{\hat{z}}}

\newcommand{\eg}{e.g.}
\newcommand{\ie}{i.e.}

\begin{document}

\title{ \Large Fast Magnetic Reconnection: Secondary Tearing Instability and Role of the Hall Term}

\shorttitle{Fast Magnetic Reconnection: Secondary Tearing Instability and Role of the Hall Term}
\shortauthors{E. Papini, S. Landi, L. Del Zanna}

%\correspondingauthor{Emanuele Papini}
\email{papini@arcetri.astro.it}

\author{\large E. Papini}
\affiliation{Dipartimento di Fisica e Astronomia, Universit\`a degli Studi di Firenze, Italy}

\author{\large S. Landi}
\affiliation{Dipartimento di Fisica e Astronomia, Universit\`a degli Studi di Firenze, Italy}
\affiliation{INAF - Osservatorio Astrofisico di Arcetri, Firenze, Italy}

\author{\large L. Del Zanna}
\affiliation{Dipartimento di Fisica e Astronomia, Universit\`a degli Studi di Firenze, Italy}
\affiliation{INAF - Osservatorio Astrofisico di Arcetri, Firenze, Italy}
\affiliation{INFN - Sezione di Firenze, Italy}

\begin{abstract}

Magnetic reconnection provides the primary source for explosive energy release, plasma heating and particle acceleration in many astrophysical environments. The last years witnessed a revival of interest in the MHD tearing instability as a driver for efficient reconnection. It has been established that, provided the current sheet aspect ratio becomes small enough ($a/L \sim S^{-1/3}$ for a given Lundquist number $S\gg 1$), reconnection occurs on ideal Alfv\'en timescales and becomes independent of $S$. Here we investigate, by means of two-dimensional simulations, the \emph{ideal} tearing instability in \edit1{both the MHD and} the Hall-MHD regime, which is appropriate when the width of the resistive layer $\delta$ becomes comparable to the ion inertial length $d_i$. Moreover, we study in detail the spontaneous development and reconnection of secondary current sheets, which for high $S$ naturally adjust to the ideal aspect ratio and hence their evolution proceeds very rapidly. For moderate low $S$, the aspect ratio tends to the Sweet-Parker scaling ($a/L \sim S^{-1/2}$). When the Hall term is included, the reconnection rate of this secondary nonlinear phase is enhanced and, depending on the ratio $d_i/\delta$, can be twice with respect to the pure MHD case, and up to ten times larger than the linear phase. Therefore, the evolution of the tearing instability in thin current sheets in the Hall-MHD regime naturally leads to an explosive disruption of the reconnecting site and to energy release on super-Alfv\'enic timescales, as required to explain \edit1{space and} astrophysical observations.
\end{abstract}


\section{Introduction}

The rapid conversion of magnetic energy into heat and particle acceleration is often \edit1{encountered in laboratory, space, and astrophysical environments. It is typically observed, e.g., in solar flares and coronal mass ejections \citep{Priest:2002, Aschwanden:2002}, and in the Earth's magnetosheath \citep{Eastwood:2018}, as well as in extreme astrophysical environments such as magnetars \citep{Lyutikov:2006}, jet and accretion disk systems \citep{Romanova:1992}, gamma-ray bursts \citep{Drenkhahn:2002}, pulsar winds \citep{Kirk:2003} and their nebulae \citep{Cerutti:2014}. 
}

On macroscopic scales, magnetized astrophysical plasmas are invariably modeled by using the MHD approximation, with a finite conductivity to be employed in Ohm's law.
However, in astrophysical systems the magnetic diffusivity $\eta$ is so small that the diffusion time  $\tau_D=L^2/\eta$ is incomparably longer than the (ideal) dynamical time scale $\tau_A=L/c_A$ required to explain such phenomena (here $L$ is the macroscopic length scale and $c_A$ the Alfv\'en speed).
The  presence of localized strong current sheets can speed up the magnetic annihilation by the mechanism of reconnection.     
Unfortunately, the classical MHD mechanisms for reconnection, namely the non-linear steady-state model by Sweet and Parker \citep{Parker:1957,Sweet:1958}, SP from now on, and the linear tearing instability \citep{Furth:1963} both predict a very inefficient reconnection rate, and the search for efficient reconnection has steadily moved from macroscopic MHD to kinetic regimes \citep[e.g.][]{Yamada:2010}.

The steady-state SP model for incompressible magnetic reconnection driven by a constant velocity inflow $v_{in}$ in a current sheet of length $L$ and width $a$, predicts a reconnection time which increases with the Lundquist number $S = \tau_D/\tau_A\gg 1$ as
\begin{gather}
 \tau_{SP}  = \tau_A ({c_A}/{v_\mathrm{in}}) = \tau_A ({L}/{a}) =  \tau_A \, S^{1/2} = \sqrt{\tau_A\tau_D},
\end{gather}
far too slow to explain the observed flare-like events, given that for astrophysical plasmas the usual estimate is $S\sim 10^{12}$ (note that the SP model also implies that the scaling for the aspect ratio of the current sheet is $L/a = S^{1/2}$).
On the other hand, current sheets are known to be locally prone to the  linear tearing instability, which leads to the formation of X-points and magnetic islands (also called \emph{plasmoids}) during the reconnection process. The $e$-folding time $\tau_t$ of the fastest growing mode, calculated by using the current sheet half thickness $a$ as characteristic length, reads 
\begin{gather}
 \tau_t  \sim {\tau_A}_a\, S_a^{1/2}  = (a/L)^{3/2} \tau_A \, S^{1/2} ,
 \label{eq:tearing_gamma}
\end{gather}
where ${\tau_A}_a = a/{c_A}$ and $S_a={a c_A}/{\eta}$. As far as $a$ is of the order of the macroscopic scale and $S\gg 1$, the timescale is again too large \edit1{\citep[analogous derivations of Eq. (\ref{eq:tearing_gamma}) can be found in][]{Bhattacharjee:2009,Pucci:2014}}.

\edit1{As hinted by the pioneering work of \citet{Biskamp:1986}, and later further developed \citep{Tajima:2002,Loureiro:2007,Lapenta:2008,Samtaney:2009,Bhattacharjee:2009,Cassak:2009}}, it has been recognized that, for high Lunquist numbers $S~\gtrsim~10^4$, a SP like current sheet undergoes tearing instability which, once measured on the relevant scale $L$, is very fast. 
Indeed, Eq.~(\ref{eq:tearing_gamma}) predicts a super-Alfv\'enic linear growth rate for the tearing instability $\gamma\sim 1/\tau_t > 1/\tau_A$, and even increasing with $S$ as $\gamma \tau_A \sim S^{1/4}$. This implies a very efficient reconnection and an explosive nonlinear production of a chain of fast moving, merging plasmoids (the so-called \emph{plasmoid instability}). This result is of course paradoxical, since the ideal MHD case (where reconnection is forbidden) cannot be retrieved for $S\to\infty$, and the only possibility to resolve this puzzle is that the SP current sheet cannot form in any dynamical thinning process \citep{Pucci:2014,Tenerani:2016,Uzdensky:2016,Landi:2017}. 

Indeed, by analyzing the characteristic timescales involved in the dynamic evolution of a \edit2{forming current sheet \citep[see, e.g.][and references therein]{Shibata:2001,Biskamp:2005,Papini:2018},} whenever the condition
\begin{gather}
 v_\mathrm{in}/c_A = a/L \le S^{-1/3}
 \label{eq:threshold}
\end{gather}
holds, the tearing instability evolves on super-ideal timescales and the sheet is disrupted. Therefore the SP configuration, even thinner than this critical threshold, can never form. 
We note that a criterion similar to Eq. (\ref{eq:threshold}) was found to  hold in the formation of secondary tearing instabilities in a plasmoid-induced reconnection model \citep{Shibata:2001}.
If one \edit1{considers a current sheet with a Harris profile and} assumes that the inverse aspect ratio scales with $S$ as $a/L = S^{-\alpha}$, the growth rate of the most unstable mode is
\begin{equation}
\gamma\tau_A \simeq 0.63 \, S^{(3\alpha - 1)/2},
\end{equation}
where the numerical factor arises from the detailed analysis of the tearing instability \edit1{(this factor changes if one considers different equilibria, but the scaling with S holds)}. Notice that the SP case is correctly retrieved for $\alpha=1/2$. The linear phase of the tearing instability for the critical case $a/L = S^{-1/3}$, named \emph{ideal} tearing instability, was first examined analytically by \citet{Pucci:2014}, who calculated the eigenmodes and found that the growth rate of the fastest reconnecting mode indeed tends asymptotically (for $S\to\infty$) to 
\begin{gather}
\gamma\tau_A \simeq 0.63,
\label{eq:ideal}
\end{gather}
that means reconnection on the macroscopic Alfv\'enic timescales. This result has been also retrieved and extended to the nonlinear regime using numerical simulations \citep{Landi:2015,DelZanna:2016a,Landi:2017}.

Related works have analyzed the evolution during the collapse of a current sheet \citep{Tenerani:2015}, the dependence on viscosity \citep{Tenerani:2015a} and on the equilibrium profile \citep{Pucci:2018}, the inclusion of electron inertia \citep{DelSarto:2016,DelSarto:2017} and the extension to the relativistic regime, in which the linear and nonlinear cases have been analyzed for the first time \citep{DelZanna:2016}.

Since the critical inverse aspect ratio can be very small, it is important to study how the ideal tearing instability is affected when \edit1{the thickness of the inner diffusive layer} approaches the ion inertial length $d_i$. 
In general, the growth rates of the tearing mode instability are known to be larger in the appropriate Hall regime \citep{Terasawa:1983,Shay:2001,Shaikhislamov:2004}. A linear analysis for the thin current sheets required for fast reconnection has been performed in the Hall-MHD regime by \citet{Pucci:2017}. The scaling for the growth rates now depends on $d_i$ as well and the growth is confirmed to be faster with respect to the MHD case. Preliminary nonlinear simulations can be found in \citet{Papini:2018}, where it is shown that secondary instabilities are also more rapidly evolving when the Hall effect is included. 

In the present paper we extend these works and study in detail, through two-dimensional MHD and Hall-MHD simulations, the development and the nonlinear stage of the tearing instability in critical current sheets with $a/L\sim S^{-1/3}$. In particular, we concentrate on the physical conditions holding at the time of the onset of secondary tearing instabilities occurring inside the main reconnecting sheet.

\section{Equations and numerical setup}

\subsection{Hall-MHD model of the tearing instability}

While the macroscopic MHD approximation is a one-fluid model, when spatial ion scales are reached the electron and the ion velocities decouple. When that happens, it is the electron velocity, defined by
\begin{equation}
\bv_e = \bv - \frac{\bJ}{e n_e}, \qquad \bJ = \frac{c}{4\pi} \nabla\times\bB
\end{equation}
($n_e$ is the numerical density of electrons, $e$ is the unsigned fundamental electrical charge, and $c$ is the speed of light), that drives magnetic evolution by entering the induction equation
\begin{equation}
 \PD{\vec{B}}{t} = \nabla\times\left ( \bv_e \times \bB\right ) + \eta \nabla^2 \bB.
\end{equation}
The full system of compressible, nonlinear Hall-MHD equations then becomes
\begin{gather}
 \PD{\rho}{t} + \DIV{(\rho \bv)} = 0
  \label{eq:continuity}
  \\
 \rho\left (\PD{}{t} + \bv\cdot\nabla \right) \bv = -\nabla P + (\CURL{\bB})\times\bB
  \\
  \left (\PD{}{t} + \bv\cdot\nabla \right) T = 
  (\Gamma \! - \! 1) \! \left [ - (\DIV{\bv})T 
  \! + \!  \frac{1}{S} \frac{|\CURL\bB|^2}{\rho} \right ]  \\
   \PD{\vec{B}}{t} = \! \nabla\!\times\!\left ( \bv \times \bB\right ) \! + \! \frac{1}{S} \nabla^2 \bB 
 -  \eta_H\nabla\!\times\! \frac{(\nabla\!\times\!\bB)\!\times\!\bB}{\rho},
  \label{eq:induction_hall_adi}
\end{gather}
where $\Gamma$ is the adiabatic index and the other variables retain their usual meaning. All quantities have been here normalized against the Alfv\'enic ones $L$, $B_0$, $\rho_0 = m_i n_0$, $c_A = B_0/\sqrt{4\pi\rho_0}$, $P_0=\rho_0 c_A^2$, $T_0 = (k_B/m_i) P_0/\rho_0$, with $m_i$ being the mass of the ions constituting the plasma. The Hall coefficient is defined as $\eta_H =d_{i}/L$, where $d_{i}$ is the reference value of the ion inertial length 
\begin{gather}
 d_i = \frac{c}{\omega_{pi}} = c \sqrt{\frac{m_i}{4\pi n_i e^2}},
\end{gather}
which depends on the plasma frequency $\omega_{pi}$ of ions and in turn on $n_i\equiv n_e$ ($=n_0$ for $d_{i}$).
\edit2{The inclusion of the Hall term is sufficient to correctly reproduce the basic properties of magnetic reconnection at ion scales, regardless of whether the magnetic field dissipation is caused by a finite resistivity or by the off-diagonal terms in the electron pressure tensor \citep{Shay:2001}. It is worth nothing, however, that kinetic effects due to the electron pressure tensor must be taken into account if one wants to accurately reproduce the geometrical and dynamical properties of the electron diffusion region \citep[see, e.g.][]{Zenitani:2011}.}

As anticipated, the Hall term is not negligible when the ion inertial length $ d_{i}$ becomes comparable to the width $\delta$ of the inner resistive layer of the tearing instability, which is smaller than the sheet's half thickness $a$. For the fastest growing mode, the inner width $\delta$ is described by the equation \citep{Biskamp:1993}
\begin{equation}
{\delta}/{a} \simeq S_a^{-3/10} \Delta'^{1/5},
\end{equation}
where $S_a$ is the Lundquist number employed in Eq.~(\ref{eq:tearing_gamma}) and $\Delta'$ is an instability parameter which depends on the configuration considered for the equilibrium magnetic field.  $\Delta'$ may depend on $ka$, which  scales as $ka\sim S_a^{1/4}$ for the fastest growing mode. For the Harris sheet configuration commonly employed in numerical works, including the present one (see the initial conditions further on)
\begin{equation}
 \Delta' = 2 \left ( {1}/{ka} - ka \right ).
 \end{equation}
Thus, at high Lundquist numbers $S_a$ and for the fastest growing mode we find $ \Delta' \sim S_a^{1/4}$ and $\delta/a \sim S_a^{-1/4}$, hence the ratio
\begin{align}
d_{i}/\delta =   \eta_H (L/a) S_a^{1/4} = \eta_H  S^{(3\alpha+1)/4}
\label{eq:di_delta}
\end{align}
is the quantity that determines whether Hall effects are important in the dynamics of reconnection \citep{Terasawa:1983}. Here the second expression is referred to the generic aspect ratio $a/L = S^{-\alpha}$ considered above, recalling that $S_a = (a/L)S$. Notice that in the critical case $\alpha=1/3$ we find
\begin{align}
 d_{i}/\delta =  \eta_H  S^{1/2}
\end{align}
and we can identify three distinct regimes: an MHD regime ($ \eta_H \ll S^{-1/2}$), where the Hall term does not play a relevant role, a mild Hall regime ($\eta_H \lesssim S^{-1/2}$), where the ion inertial length is comparable to the thickness of the inner layer, and a strong Hall regime ($\eta_H > S^{-1/2}$), where reconnection is dominated by the Hall effect and the classic theory of the tearing instability is no longer valid \citep[see also][]{Shaikhislamov:2004}. Recently, \citet{Pucci:2017}  extended the ideal tearing instability to include the Hall term in the case of a Harris current sheet in pressure equilibrium, that is an unperturbed magnetic field which is unidirectional. They found the existence of the regimes discussed above, and showed that the linear growth rate starts to increase roughly for values $d_i/\delta \sim 3$ \edit1{(note that \citet{Pucci:2017} define $P_h=d_i/\delta$)}.  However, this threshold is likely to be actually even smaller, since Hall currents may affect the subsequent nonlinear evolution, where thinner current sheets that formed between the ejected plasmoids may eventually host secondary reconnection events, as we will show later in Section~\ref{sec:secondrec}.

\subsection{Numerical setup}

In the present work we consider nonlinear simulations with initial condition for the magnetic field given by a two-dimensional force-free (FF) current sheet  configuration, centered at $x=0$ and asymptotically aligned in the $y$-direction with $|\bB| = B_0= 1$ according to the profile
\begin{gather}
 \bB_0 = \tanh(x/a) \bey + \mathrm{sech} (x/a) \bez,
 \label{eq:harris_ff}
\end{gather}
which reproduces a Harris profile for the in-plane component, but keeps a constant magnetic (and thus total) pressure by rotating the magnetic field around the $x$-axis.
Moreover, we assume homogeneous density $\rho =1$, pressure and temperature $P=T=\beta/2$, and we do not impose initial velocities ($\bv_0=0$).  The plasma beta is chosen as $\beta=0.5$, the adiabatic index is $\Gamma = 5/3$, and we investigate the case appropriate for the \emph{ideal} tearing, thus we choose the half thickness of the current sheet $a = S^{-1/3}=\eta^{1/3}$, for $L=1$. We follow the evolution of the plasma by integrating the system of Hall-MHD equations (\ref{eq:continuity} -\ref{eq:induction_hall_adi}) in a $[-L_x,L_x]\times [0,L_y]$ domain in the $xy$-plane, but retaining all components of the 3D vectors. The in-plane magnetic field is evolved through a scalar potential $\phi$ (the $z$ component of the vector potential), so that $B_x = \partial_y\phi$ and $B_y = - \partial_x\phi$, in order to preserve the solenoidal constraint for the magnetic field. 

\begin{table*}[t]
\begin{center}
 \begin{tabular}{c|rcccccc}
  Run & $S (\eta^{-1})$  & $a/L$ & $\eta_H$ & $d_{i}/\delta$ & $L_x$ & $L_y$ & $N_x \times N_y$ \\ \vspace*{-0.3cm}\\\hline 
  0L & $10^5$ & $\sim0.022$ & 0.01  & $\sim3.2$ & $\sim0.43$ & $\sim2.2\pi$ & $1024\times128$ \\
  1L & $10^5$ & $\sim0.022$ & 0     & 0   & $\sim0.43$ & $\sim2.2\pi$ & $1024\times128$ \\
  2L & $10^5$ & $\sim0.022$ & 0.002 & $\sim0.6$ & $\sim0.43$ & $\sim2.2\pi$ & $1024\times128$ \\
  3L & $10^5$ & $\sim0.022$ & 0.005 & $\sim1.6$ & $\sim0.43$ & $\sim2.2\pi$ & $1024\times128$ \\
  4L & $10^5$ & $\sim0.022$ & 0.01  & $\sim3.2$ & $\sim0.43$ & $\sim2.2\pi$ & $1024\times128$ \\
  5N & $10^5$ & $\sim0.022$ & 0     & 0   & $\sim0.43$ & $\sim2.2\pi$ & $4096\times512$ \\
  6N & $6.7\cdot 10^5$ & $\sim0.011$ & 0 & 0    & $\sim0.23$ & $\sim1.14\pi$ & $4096\times512$ \\
  7N & $8\cdot 10^5$ & $\sim0.011$ & 0      & 0 & $\sim0.22$ & $\sim1.08\pi$ & $4096\times512$ \\
  8N & $10^6$ & 0.01 & 0      & 0         & 0.2     & $\pi$ & $4096\times512$ \\
  9N & $10^6$ & 0.01 & 0.0002 & $\sim0.2$ & $0.2$ & $\pi$ & $4096\times512$ \\
  10N & $10^6$ & 0.01 & 0.0014 & $\sim1.4$ & $0.2$ & $\pi$ & $4096\times512$ \\
  11N & $10^6$ & 0.01 & 0.003  & $\sim3.0$ & $0.2$ & $\pi$ & $4096\times512$ \\
  12N & $2\cdot 10^6$ & $\sim0.008$ & 0      & 0  & $\sim0.16$ & $\sim0.79\pi$ & $4096\times512$ \\
  13N & $5\cdot 10^6$ & $\sim0.006$ & 0      & 0  & $\sim0.12$ & $\sim0.58\pi$ & $4096\times512$ \\
  14N & $10^7$ & $\sim0.0046$ & 0      & 0   & $\sim0.093$  & $\sim0.46\pi$ & $4096\times512$ \\
  \edit1{15N} & $10^8$ & $\sim0.0022$ & 0      & 0   & $\sim0.043$  & $\sim0.22\pi$ & $4096\times512$ \\
 \end{tabular}
\end{center}
 \caption{Physical and numerical parameters of all the simulations performed in this work for linear (L) and nonlinear (N) runs. From left to right: Lundquist number $S$, aspect ratio $a/L$, Hall coefficient $\eta_H=d_i/L$, ratio of the ion inertial length $d_i$ with respect to the inner resistive layer $\delta$, domain size across ($L_x$) and along ($L_y$) the current sheet, and grid size ($N_x \times N_y$). Run 0L starts from a pressure equilibrium (PE) configuration, all the other runs use a force-free (FF) equilibrium.}
\label{tab:sim_set}
\end{table*}
At the beginning of the simulation, the tearing instability is triggered  by in-plane magnetic perturbations localized inside the current sheet. In terms of the scalar potential these perturbations take the form
\begin{gather}
 \phi = \varepsilon \, {\rm{sech}} (x/a) \sum_{n=1}^{N}\cos(k_n y +\varphi_n), 
\end{gather}
where $k_n=2\pi n/L_y$ and $\varphi_n$ is a random phase different for each value of $n$. We choose $N=10$ and  $\varepsilon = 10^{-6}$ (the overall amplitude of the perturbed magnetic field is $\sim 10^{-4}$). The value of $L_y$ is chosen such that the lowest wavenumber resolved for the tearing instability is $ka = 2\pi a/L_y = 0.02 $. This value is more than sufficient to capture the fastest growing mode of the tearing instability for the values of $S$ considered in this work. In the $x$-direction we set $L_x=20a$ to have boundaries sufficiently far from the reconnection region while retaining the high resolution required inside the current sheet.   

The Hall-MHD equations are numerically solved by means of the same MHD code we used in \citet{Landi:2015}, modified to include the Hall term. Spatial derivatives are calculated using Fourier decomposition along the periodic direction and a fourth-order compact scheme \citep{Lele:1992} across the current sheet. Time integration is performed with a third-order Runge-Kutta scheme, taking into consideration the effect of the Hall term in the definition of the timestep. Boundary conditions are periodic along $y$ (only integer numbers of wavelengths are thus allowed) and of free outflow at $x=\pm L_x$, using the method of projected characteristics \citep{Poinsot:1992,DelZanna:2001,Landi:2005}. Unless differently specified, the employed grid consists of $N_x \times N_y = 4096\times512$ points, which allows us to resolve secondary reconnection events in both the $x$- and the $y$-directions.
Table \ref{tab:sim_set} report the full set of simulations used in this work.

\section{Linear phase}
\label{sec:linear_phase}
We now describe the evolution of the linear tearing instability in the case of the ideal limit $a/L = S^{-1/3}$. The results of this section confirm the findings of \citet{Pucci:2017}, where pressure equilibrium was imposed, and extend them to the force-free case, more appropriate for magnetically dominated systems, employed here and in many other works.  The initial magnetic equilibrium is here considered in the general form $\bB_0=(0,B_{0y}(x),B_{0z}(x))$.
The governing equations of the linear tearing instability read
\begin{align}
\label{eq:tearing_h0}
 \gamma (v_x'' - k^2 v_x) = & ik \left [ B_{0y} (b_x'' -k^2b_x) - B_{0y}''b_x\right ], \\
 \gamma b_x = & ikB_{0y} v_x + \frac{1}{S} (b_x'' - k^2b_x) 
 \nonumber \\
    & + \frac{\eta_H}{\rho_0} \left ( k^2 B_{0y}b_z - ikB_{0z}'b_x  \right ), \\
  \gamma b_z = &ikB_{0y} v_z - B_{0z}' v_x  + \frac{1}{S} (b_z'' - k^2 b_z) \nonumber\\
    &+ \frac{\eta_H}{\rho_0} \left [ B_{0y}(b_x'' - k^2 b_x) - B_{0y}''b_x  \right ],
\\
        \gamma v_z = &ik B_{0y} b_z + b_x B_{0z}',
        \label{eq:tearing_h1}
\end{align}
and for a given set of parameters $k, S,$ and $\eta_H$, the above system of equations constitute a 12th-order two-points eigenvalue problem. Here $v_x,v_z,b_x,$ and $b_z$ are the (complex) eigenfunctions of the $x$ and the $z$ components of velocity and magnetic field respectively, and the apex denotes differentiation with respect to $x$. Each eigenmode perturbation, \eg~$b_x$, has the form $b_x(x,y,t)= b_x(x) e^{\gamma t + iky}$, where $k$ is the wavenumber associated to the perturbation in the $y$-direction and the eigenvalue $\gamma$ is the corresponding linear growth rate. In the following we will assume the same settings employed in our numerical simulations, that is, a FF equilibrium (see Eq.~(\ref{eq:harris_ff})).

The above set of equations holds also for the case of pressure equilibrium (PE), in which all terms with $B_{0z}'$ vanish so that the equations simplify to the analogous ones employed in \citet{Pucci:2017}: the eigenvalue problem reduces to 6th-order and, in analogy with the MHD case, the eigenfunctions for the magnetic field and for the velocity are purely real and imaginary, respectively. Moreover, as for the MHD classical tearing instability, we see that in a PE configuration the presence of a constant guide field is ineffective. In the FF case, however, three additional terms involving $B_{0z}'$ appear in the equations, and these will lead to different results, especially on the parity of some of the eigenfunctions.

\begin{figure}[t]
\includegraphics[width=\columnwidth]{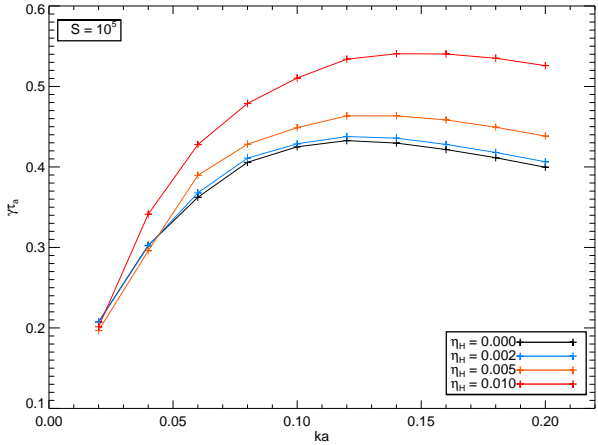}
\caption{
Growth rates $\gamma\tau_A$ vs. $ka$ in the linear phase of Hall-MHD simulations with $S=10^5$ and different values of $\eta_H$ (Run 1L-4L). The first ten wavenumbers with $ka$ from $0.02$ to $0.2$ have been excited.
}
 \label{fig:gammatauvska10e5}
\end{figure}

\begin{figure*}
\includegraphics[width=\textwidth]{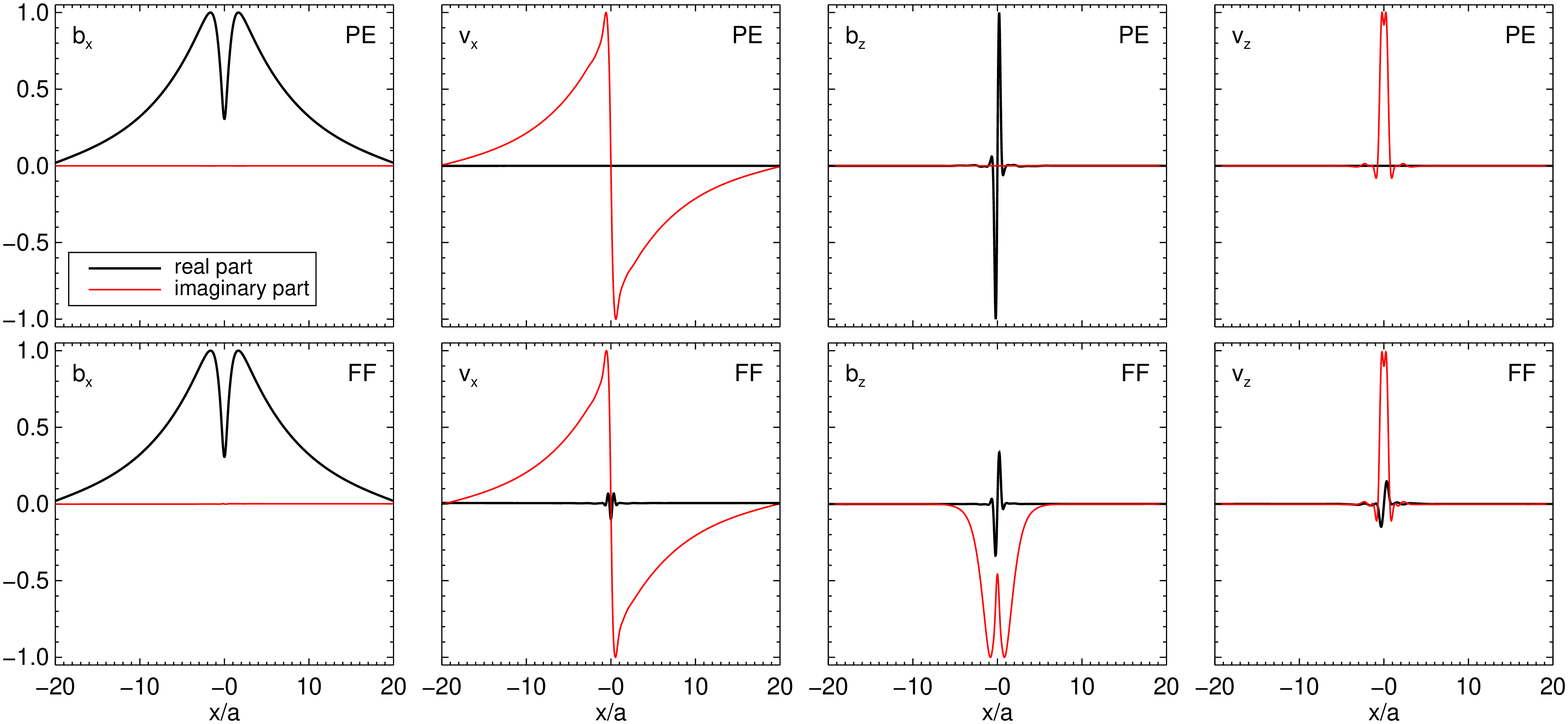}
\caption{Eigenfunctions of the linear Hall tearing instability $b_x,v_x,b_z,v_z$ for a mode with $ka=0.14$, in the pressure equilibrium (PE, top panels, Run 0L) and in the force-free (FF, bottom panels, Run 4L) case, as extracted from two Hall-MHD simulations with $S=10^5$ and $\eta_H=0.01$. Solid black and red lines denote, respectively, real and imaginary part of each eigenfunction. Each (complex) eigenfunction is normalized with respect to its maximum modulus.}
 \label{fig:eigenfunctions}
\end{figure*}

In all simulations, a linear tearing instability develops at the beginning and with the same qualitative behavior. There are, however, some quantitative differences due to the presence of the Hall term. 
To highlight them, we calculated the linear dispersion relation for four simulations with the same value of $S=10^5$ but different values of $\eta_H=0$ (MHD case), $ 0.002, 0.005,\text{ and } 0.01$, corresponding to $d_i/\delta= 0, 0.6, 1.6, \text{and } 3.2$, respectively (see Run 1L-4L of Table \ref{tab:sim_set}).
The linear growth rate $\gamma$  has been calculated by taking, at each time of the linear phase, the modulus of the Fourier transform along the $y$-direction of the average along the $x$-direction of $B_x$, since its eigenfunctions are even with respect to $x$ and since  $B_x$ has no equilibrium component. An exponential fit has been then performed, separately for each Fourier component, to obtain the linear growth rate $\gamma$. These dispersion relations are reported in Fig.~\ref{fig:gammatauvska10e5} and have a similar shape in all cases. In general, for larger $\eta_H$ values the corresponding curve yields larger values of $\gamma$. More precisely, the linear growth rate of each mode increases when $d_i$ exceeds the thickness of the inner resistive layer $\delta$, up to about $25\%$ more than the MHD case for $ka=0.14$ and $\eta_H=0.01$. This is in qualitative and quantitative agreement with \citet{Pucci:2017}, even though the initial equilibrium here is different and therefore the linear evolution may also be different, due to the additional terms present in Eqs.~(\ref{eq:tearing_h0}-\ref{eq:tearing_h1}). Notice that the results are lower than expected theoretically. For instance, in the MHD case we observe a maximum rate roughly $20\%$ lower than the value predicted by theory. This difference was also encountered in \citet{DelZanna:2016} and it is due to the diffusion of the initial equilibrium during the evolution. More accurate results were obtained in \citet{Landi:2015}, where this effect was properly treated.

Differences between the FF and the PE equilibrium arise in the eigenfunctions, shown in Fig.~\ref{fig:eigenfunctions}. The eigenfunctions have been obtained by using a linearized version of our Hall-MHD code \citep{Landi:2005} and are quantitatively and qualitatively similar to the ones observed in the linear phase of the fully nonlinear simulations.
Indeed, the eigenfunctions $b_x$, $v_x$, $b_z$ and $v_z$ extracted by a numerical simulation with a PE configuration are even, odd, odd, and even, respectively, as in \citet{Pucci:2017}. Moreover, $v_x$ and $v_z$ are purely imaginary, while $b_x$ and $b_z$ are real. The parity relations can be written as 
\begin{gather}
 1=\rP(b_x^\rR) = -\rP(b_z^\rR) = - \rP (v_x^\rI) = \rP (v_z^\rI),
 \label{eq:parity_PE}
\end{gather}
where $\rP$ denotes the parity operator, whereas the '$\rR$' and '$\rI$' superscripts indicate the real and the imaginary part, respectively. In the FF configuration, the eigenfunctions are complex. The parity relations (\ref{eq:parity_PE}) hold also in the FF case, complemented by the relations 
\begin{gather}
 1=-\rP(b_x^\rI) = \rP(b_z^\rI) =  \rP (v_x^\rR) = -\rP (v_z^\rR)
 \label{eq:parity_FF}
\end{gather}
for the imaginary part of $b_x$ and $b_z$ and for the real part of $v_x$ and $v_z$.

\section{Nonlinear phase: General Properties}

\begin{figure}
 \includegraphics[width=1.03\columnwidth]{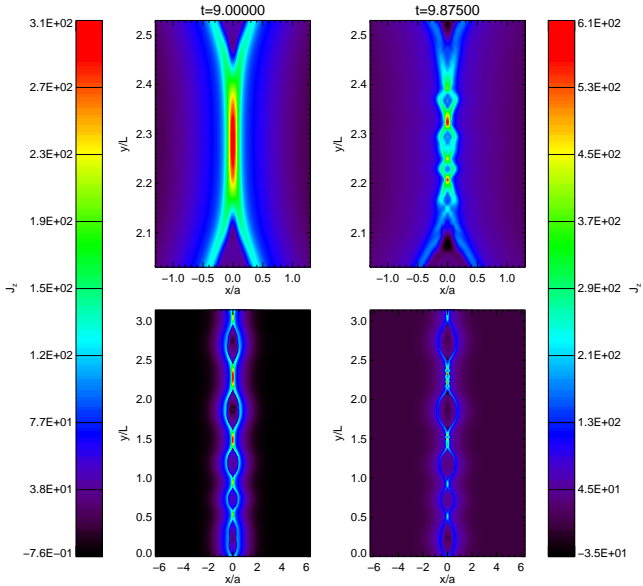}
 \caption{
Color-filled contour plots of $J_z$ for a simulation in the MHD case with $S=10^6$ (Run 8N) at two different times: just before a tearing instability is triggered in the secondary current sheets (left panels), and during the secondary reconnection events (right panels, notice the different color scale). The top panels show a zoomed region around $y\simeq 2.3\,L$, to single out the dynamics of a secondary sheet.
}
 \label{fig:jzcontour}
\end{figure}

We now focus on the nonlinear phase of the instability, and consider simulations with a higher Lundquist number, $S=10^6$, so that the settings for the \emph{ideal} tearing lead to a half thickness $a = S^{-1/3} L = 0.01\, L$. We firstly illustrate the general properties by discussing the results of the purely MHD case (Run 8N of Table \ref{tab:sim_set}), while differences due to the Hall effects will be discussed in \edit1{Section \ref{sec:second_hall}}.

In all simulations, as the linear phase evolves, the amplitudes of the perturbations increase exponentially, until the tearing instability saturates and the nonlinear phase begins, as shown in Fig.~\ref{fig:jzcontour}. There, two snapshots of the MHD simulation are taken at the beginning of the nonlinear phase, and a colored contour plot of $J_z$ is shown. At time $t=9~\tau_A$ (bottom left panel) the plasmoids have a size comparable to the thickness of the current sheet, and some of them have already merged. Among the plasmoids we also observe that secondary current sheets have formed, with a thickness of roughly one tenth of the original thickness. One of them is shown in the top left panel of the same figure, by zooming in the region centered at $y\simeq 2.3\,L$.
The subsequent evolution is characterized by the coalescence and nonlinear growth of the plasmoids, but the most important feature is the onset of secondary reconnection events in the newly formed current sheets, which then drive the dynamics and eventually lead to the disruption of the whole system.
These secondary tearing instabilities are indeed very fast, since already at time $t=9.875~\tau_A$, in less than one macroscopic Alfv\'en time, they are fully developed  (see the right panels). A more detailed analysis of the evolution of these secondary current sheets is performed in section~\ref{sec:secondrec}.

\begin{figure}
 \includegraphics[width=\columnwidth,trim={0 1.4cm 0 0},clip]{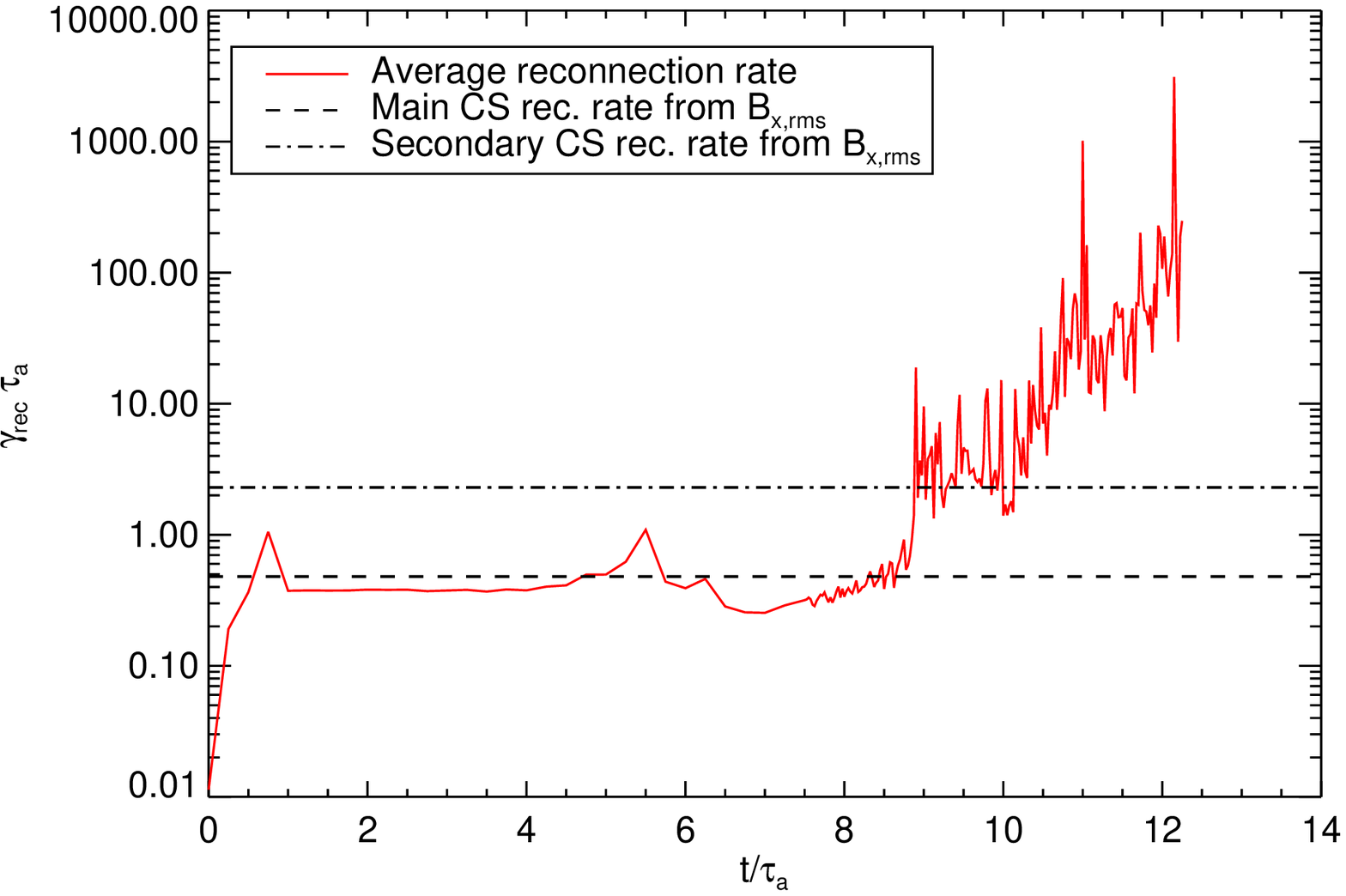}
 \includegraphics[width=\columnwidth,trim={0 1.4cm 0 0},clip]{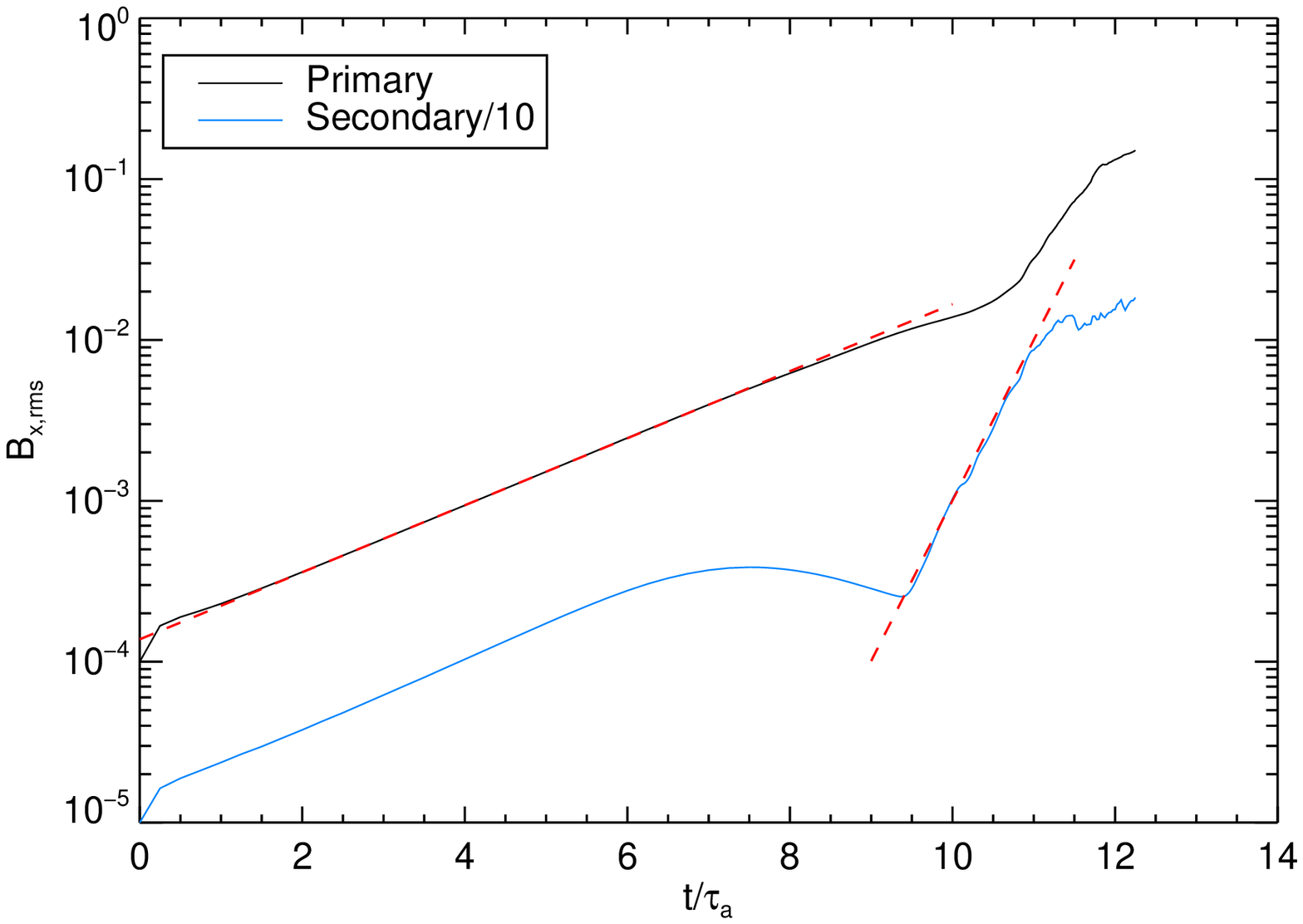}
 \caption{
Top panel: averaged reconnection rate $\gamma_\mathrm{rec}$ vs. time for the MHD case with $S=10^6$ (Run 8N). Horizontal lines are the linear growth rates as calculated with exponential fits of $B_{x,\mathrm{rms}}$ (see below). Bottom panel: $B_{x,\mathrm{rms}}$ vs. time for the primary sheet (black solid line) and for the secondary one shown in the top panels of Figure~\ref{fig:jzcontour} (blue solid line, values have been divided by a factor 10). Red dashed lines indicate the exponential fits for growth rates.
}
\label{fig:recrates1e6h0}
\end{figure}

Let us now provide a more quantitative support to the above statements. We define the averaged reconnection rate as the quantity
\begin{gather}
 \gamma_\mathrm{rec} = \frac{1}{N}\sum_{i=1}^N \frac{1}{\Phi_i}\TD{\Phi_i}{t},
\end{gather}
obtained by taking, at each time, the logarithmic time derivative of the reconnected flux $\Phi_i$ between the $i$-th pair of X- and O-points and then averaging over the number of pairs, $N$, in the main current sheet. Here, $\Phi_i$ is the difference between the scalar potential $\phi$ at the O- and the X-point in the $i$-th pair \citep[for more details and applications to simulations of plasma turbulence see the Appendix of][]{Papini:2019}.
The top panel of Figure~\ref{fig:recrates1e6h0} shows $\gamma_\mathrm{rec}$ for our MHD reference run. As we can see, after the initial perturbations have rearranged to select the fastest tearing eigenmodes, $\gamma_\mathrm{rec}$ reaches a plateau with a value $\gamma_\mathrm{rec}\,\tau_A\simeq 0.4$. It is also possible to identify a second, more noisy, plateau between $t\simeq 9\,\tau_A$ and  $t\simeq 10\,\tau_A$, roughly the temporal range selected in Fig.~\ref{fig:jzcontour}, that we interpret as the average reconnection rate of the secondary current sheets. 

In order to support this conclusion, we estimated the reconnection rates by performing an exponential fit of the root-mean-square (rms) value of the $x$-component of the magnetic field, that we name $B_{x,\text{rms}}$, which is a good proxy of the reconnection rate. The bottom panel of Fig.~\ref{fig:recrates1e6h0} shows this quantity as a function of time. The black curve denotes the primary current sheet, while the blue curve has been calculated by restricting to the secondary current sheet (the values are lowered by a factor of 10 for ease of presentation). In the latter case we notice a steepening at $t\simeq 9.5~\tau_A$, a clear signature of the secondary tearing instability. The horizontal dashed and dot-dashed lines in the top panel, with values $\gamma_\mathrm{rec}\,\tau_A = 0.48$ and $2.30$, respectively, correspond to the exponential fits indicated by the red dashed lines in the bottom panel and nicely match the two plateaux we identified. Indeed, the measured reconnection rate of the secondary current sheet is strongly super-Alfv\'enic. 

In the final stage of the evolution the secondary reconnection events drive the dynamics, with new plasmoids being ejected by super-Alfv\'enic outflows and feeding the huge plasmoids generated by the first reconnection event. Eventually, the whole current sheet is disrupted.

\section{Secondary \emph{ideal} tearing instabilities}
\label{sec:secondrec}

The study of the formation of secondary current sheets and their disruption by the onset of secondary tearing instabilities is obviously very important, since the observed secondary reconnection events have super-Alfv\'enic growth rates and drive the final evolution of the whole system. 
The aim of this section is to further characterize the spontaneously formed secondary current sheets before their evolution toward the final breakup.
We will show that, in a dynamically evolving plasma environment, it is possible to form current sheets in local (and provisional) equilibrium which then evolve on small and local temporal and spatial scales in an explosive way. It is worth nothing that such substructures are \edit1{enclosed} within a global structure (the primary current sheet) which, on the contrary, is out of equilibrium and has already evolved in a highly turbulent state. 
Therefore, in a broader context results of this section have potential implications for what concerns the dynamics of turbulent systems.

Although morphologically different, the behavior of the evolution of all Hall-MHD runs is qualitatively similar to that of the MHD ones, the growth rates being larger and the final stage more violent for $\eta_H \neq 0$ \edit1{(see Section \ref{sec:second_hall})}, thus in the present section we focus only on purely MHD simulations.

\begin{figure*}
 \includegraphics[width=0.5\textwidth]{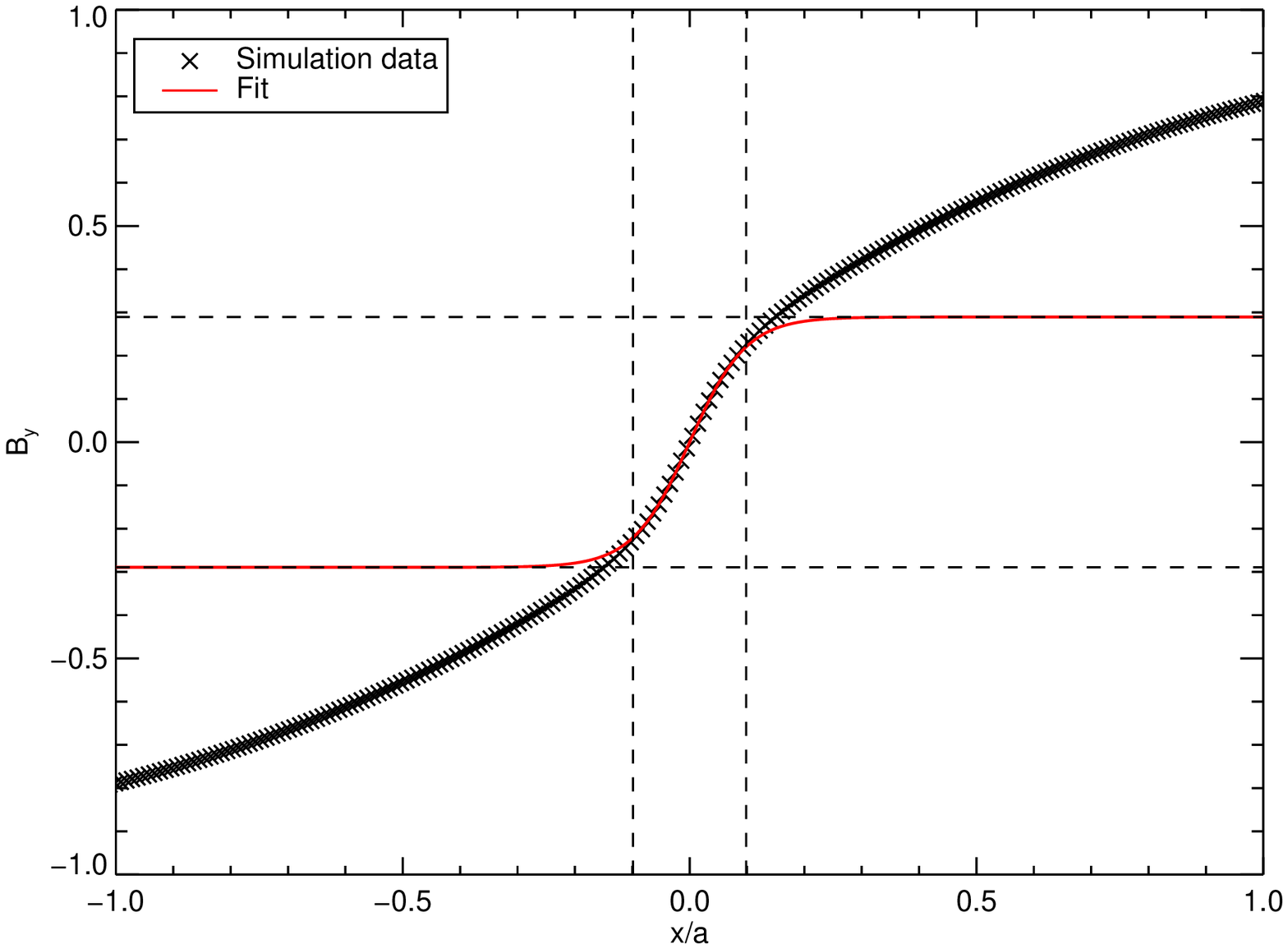}
 \includegraphics[width=0.5\textwidth]{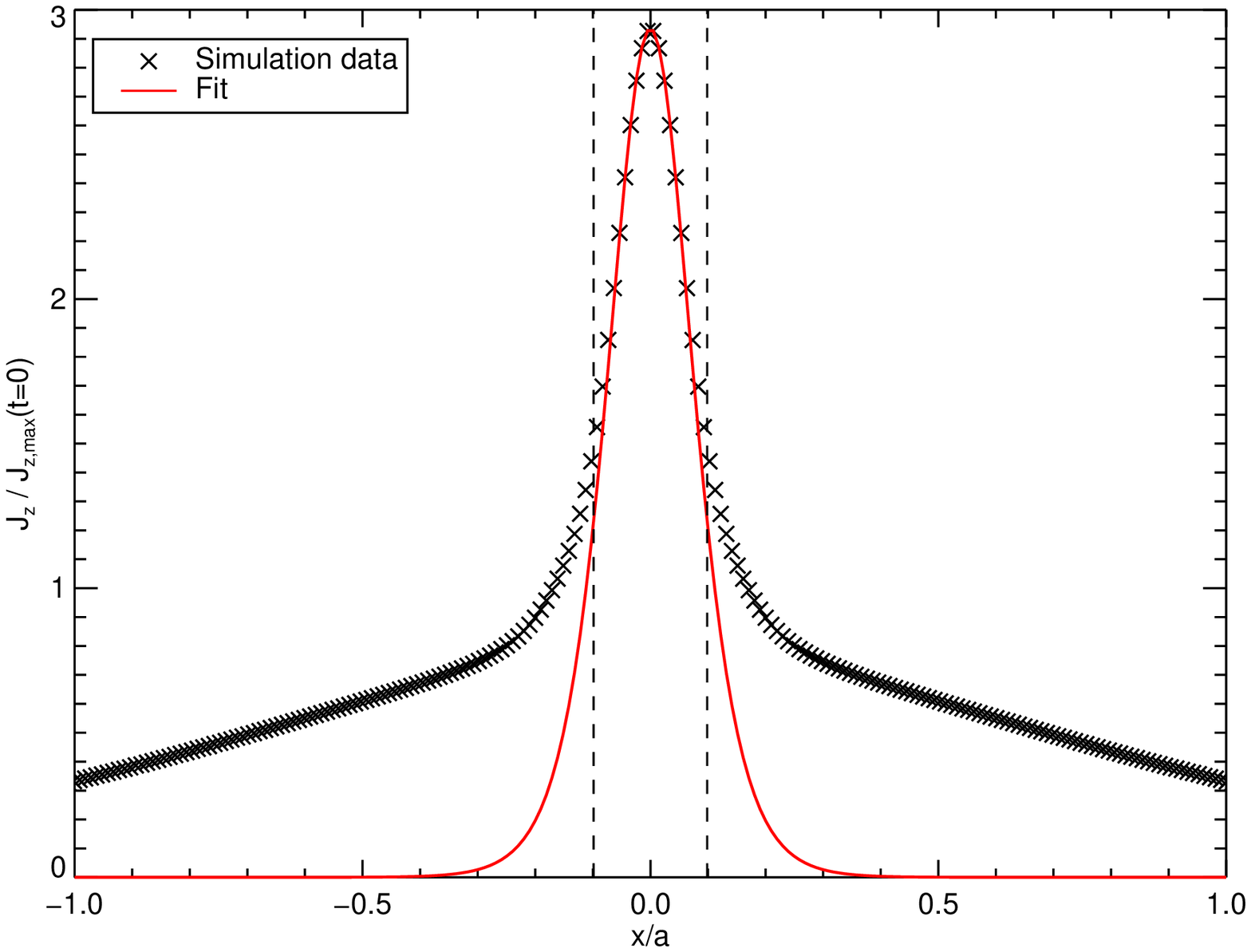}
 \caption{
 Fit of the secondary current sheet  shown in the top-left panel of Fig.~\ref{fig:jzcontour}. Horizontal and vertical dashed lines denote the fitted value of $\mathrm{B}_{0}^*$ and of the half thickness $a^*$ of the secondary current sheet, respectively, using the form in Eq.~(\ref{eq:tanh_fit}).}
\label{fig:fit_secondary}
\end{figure*}

\begin{figure}
 \includegraphics[width=\columnwidth]{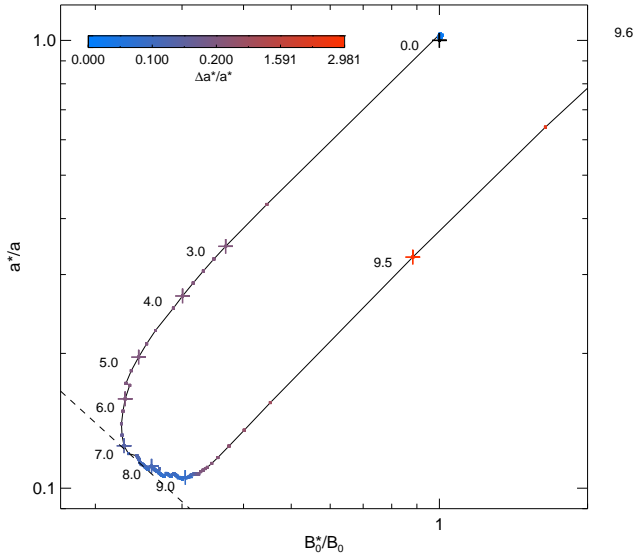}
 \caption{
 Evolution of $a^*/a$ vs. $\mathrm{B}_0^* / \mathrm{B}_0$ for the secondary current sheet  shown in the top-left panel of Fig.~\ref{fig:jzcontour}. Crosses and numbers mark specific simulation times in units of the global Alfv\'en time $\tau_A$. The dashed line denotes a reference constant value of $S_a^*$. This phase corresponds to the smallest values of $\Delta a^*/a^*$ (blue color). Later we find $\Delta a^*/a^* > 1$ (red color), indicating the final reconnection stage.
 }
\label{fig:scaling_time}
\end{figure}

We have already shown that secondary current sheets naturally form between consecutive plasmoids at the beginning of the nonlinear phase, with an approximate thickness which is $10\%$ to $15\%$ of the initial thickness $a$ (see Fig.~\ref{fig:jzcontour}). These current sheets further thin on a timescale of a couple of Alfv\'en times, until they reach a critical aspect ratio and become unstable to a secondary tearing instability. The formation of these secondary events is spontaneous, without any prior imposition on their equilibrium or their aspect ratio, therefore it is very interesting to characterize their evolution and the conditions under which the secondary tearing instabilities are triggered. 

To that purpose, for a given simulation, we identified the region where a secondary current sheet had formed. Then we measured the position of its center, $y_\mathrm{cs}$, and we calculated its length $2L^*$ by measuring the full-width-half-maximum of the current density profile $J_z$ along $y$ at $x=0$, after the background current of the primary sheet had been subtracted. Moreover, by assuming a standard profile of the form
\begin{gather}
\bB^*(x) = \mathrm{B}_0^* \tanh(x/a^*) \bey,
\label{eq:tanh_fit}
\end{gather}
and by performing a least square fit, we obtained the half thickness $a^*$ and the asymptotic magnetic field $\mathrm{B}_0^*$ of the secondary current sheet. Figure~\ref{fig:fit_secondary} shows an example of the fit, performed at the center of the secondary current sheet of the top-left panel of Fig.~\ref{fig:jzcontour}. The local Lundquist number is then found as $S_a^* = a^* \mathrm{B}_0^*/\eta$ (the density in the secondary sheets increases typically only by about $1\%$ of $\rho_0=1$, therefore we can safely identify $\rB_0^*$ with the local Alfv\'en speed). Notice that, as it will be discussed later, these dynamically formed current sheets are in a state of almost perfect pressure equilibrium, hence we do not expect that a $z$ component of the magnetic field is needed to balance the magnetic pressure in a force-free state. We thus deem that Eq.~(\ref{eq:tanh_fit}) represents the best shape for the magnetic field  to be used as a fit for the secondary current sheets.

\begin{figure*}[t]
\includegraphics[width=\columnwidth]{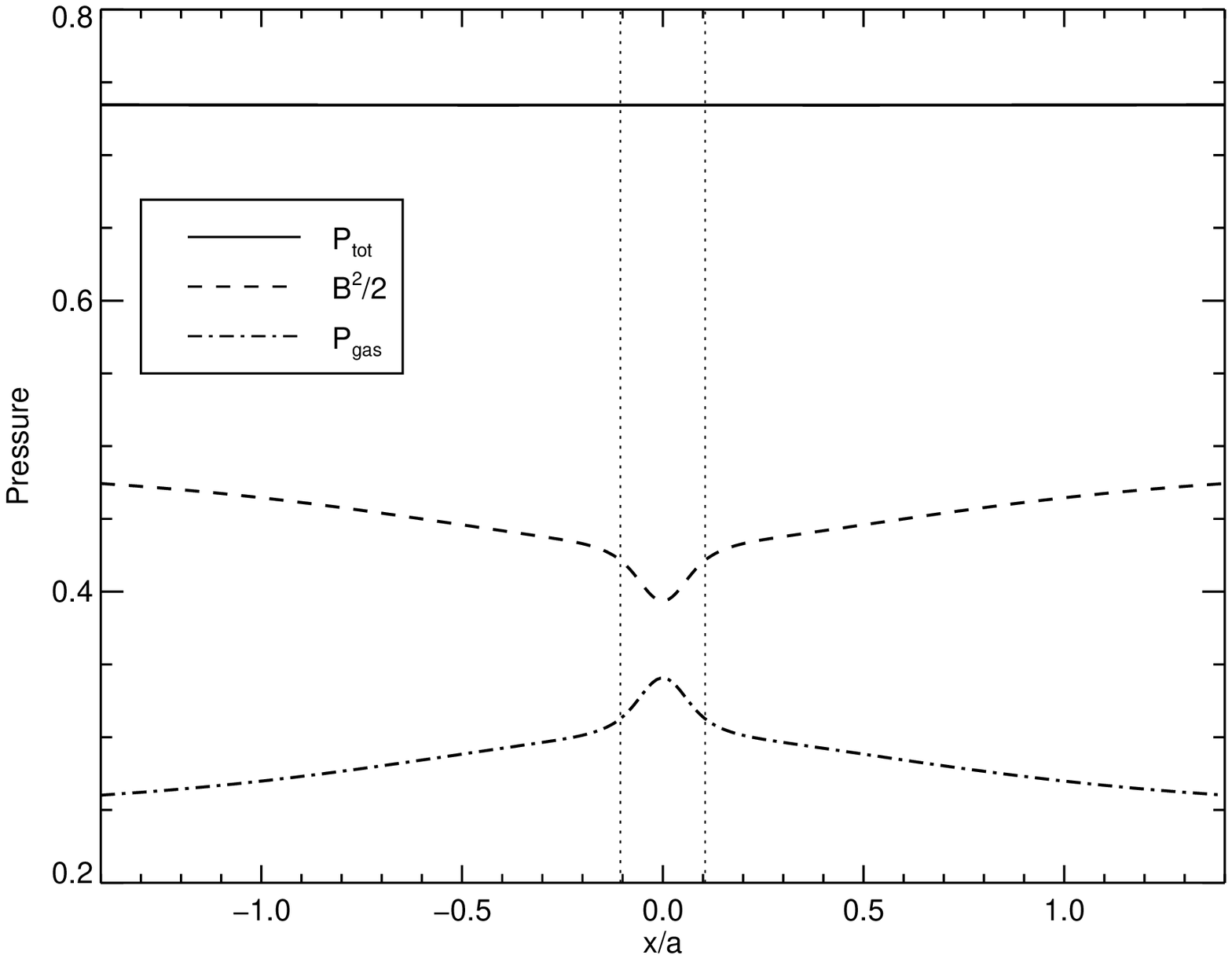}
\includegraphics[width=\columnwidth]{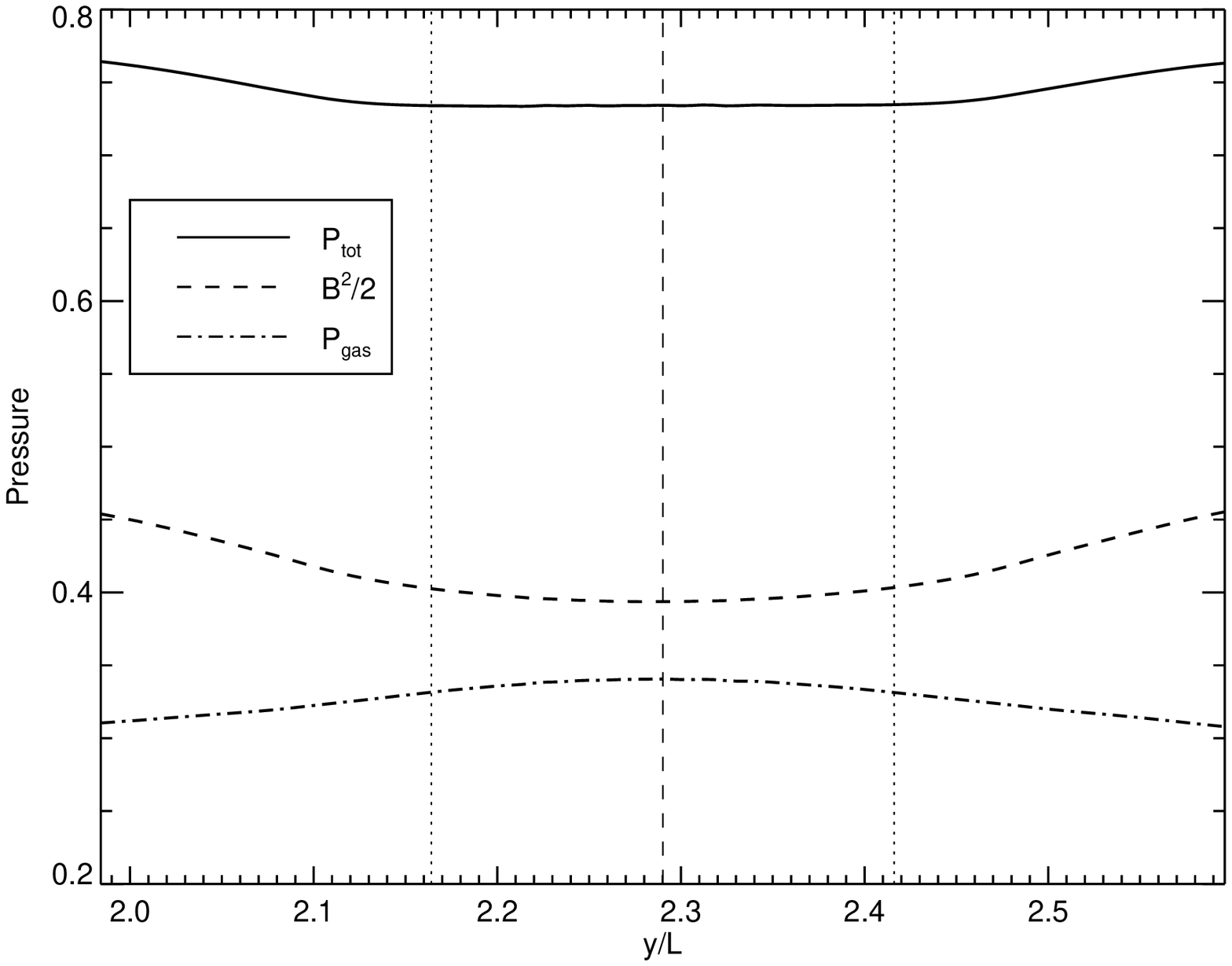}
\includegraphics[width=\columnwidth]{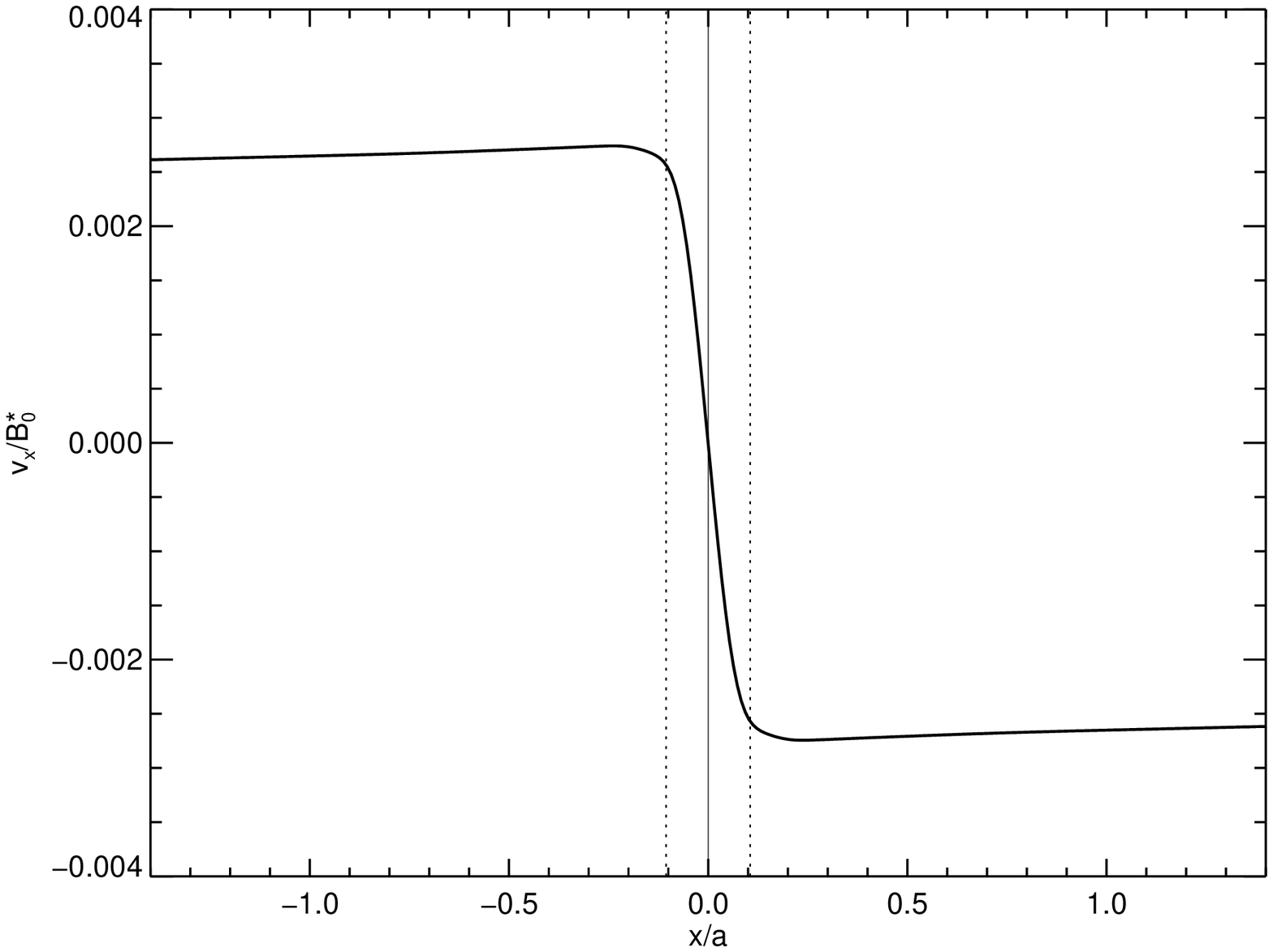}\hspace{.7 cm}
\includegraphics[width=\columnwidth]{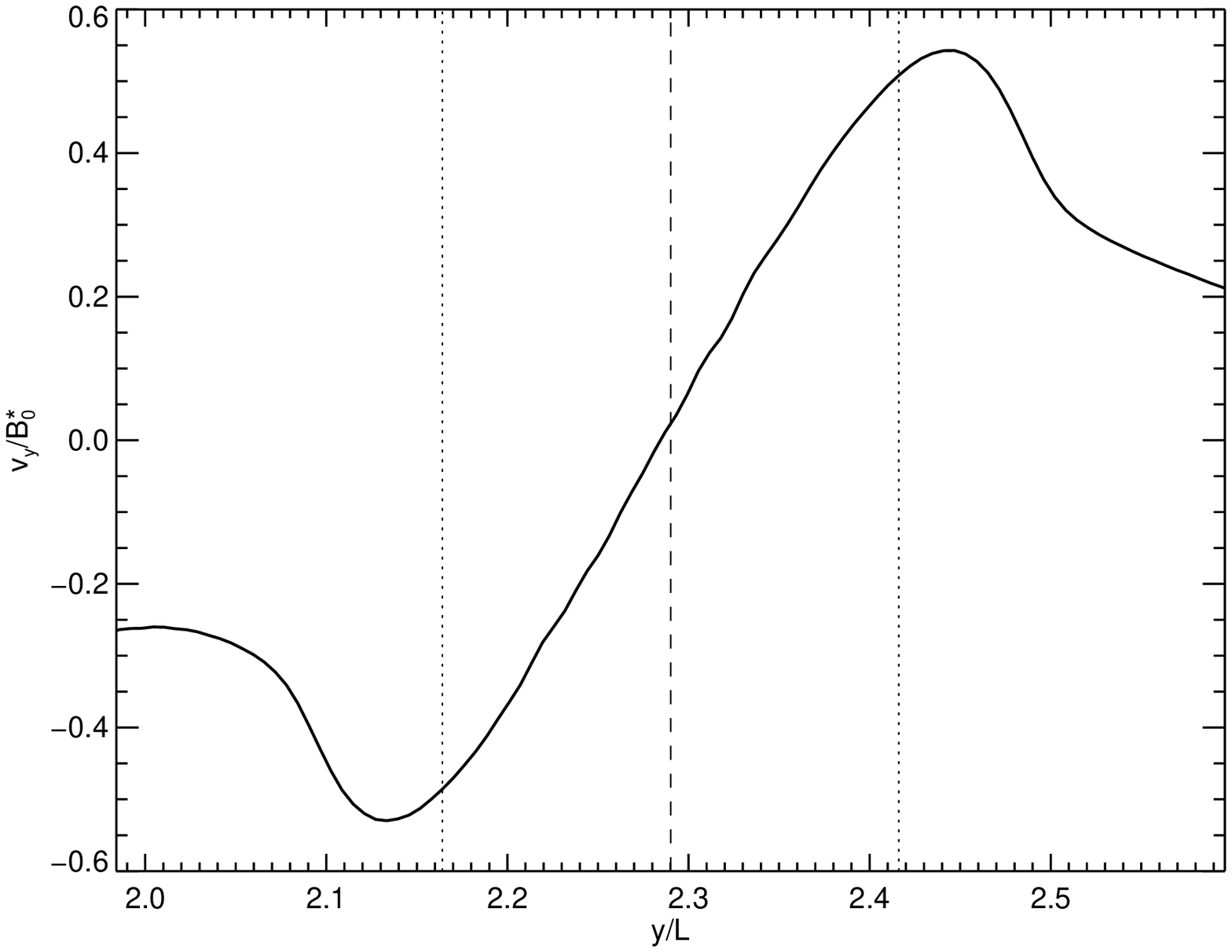}
 \caption{Profiles of the total pressure (solid line) taken at the center $y_\text{cs}$ of a secondary current sheet along the $x$-direction (top left panel) and at $x=0$ along the $y$-direction (top right panel) at the time $t=9~\tau_A$ for the MHD reference Run 8N with $S=10^6$. The dashed and dot-dashed curves are the magnetic and the gas pressure, respectively. The profiles of the inflow velocity $v_x(x,y_\text{cs})$ and of the outflow velocity $v_y(0,y)$, both normalized against the local Alfv\'en speed $c_A^*\simeq \rB_0^*$, are also shown (bottom left and right panels). Vertical dotted lines denote the thickness and the length of the secondary sheet as calculated by our fitting procedure. The vertical dashed line denotes $y_\text{cs}$.  }
 \label{fig:secondary_equilibrium}
\end{figure*}
In order to provide a statistically significant measure of $a^*$ and $\mathrm{B}_0^*$, a separate fit of the $y$-component of the magnetic field, $B_y(x,y_i)$, is performed for each grid coordinate $y_i$ in the range $[y_\mathrm{cs}-L^*/2,y_\mathrm{cs} +L^*/2]$ \edit1{(i.e. in the central half of the current sheet), in order to obtain two sets $\{a_i^*\}$ and $\{\rB_{0,i}^*\}$ of the desired quantities.
Finally, the best value and error of $a^*$ and $\mathrm{B}_0^*$ are taken as the mean and the standard deviation of the above sets.} We further note that the standard deviation $\Delta a^*$ is larger than $a^*$ when the secondary current sheet is either in its nonlinear reconnection phase or it has not formed yet. Therefore, values $\Delta a^*/a^* \ll 1$ indicate the phase in which a well defined secondary current sheet is present, while values $\Delta a^*/a^* > 1$ give a rather precise indication of when secondary reconnection events are about to disrupt it.

To track the evolution in time of the secondary current sheet, we performed the above fitting procedure for all the outputs of the simulation. Figure~\ref{fig:scaling_time} shows the evolution of $a^*$ and $\rB_0^*$ of the secondary current sheet already discussed in the MHD reference run. The value of $\Delta a^*/a^*$ is color coded, so to capture the formation of the secondary current sheet. In particular, the blue points \edit1{(e.g. at $t=8\,\tau_A$)} indicate a well defined current sheet, with $\Delta a^*/a^* \ll 1$, whereas red points either denote the thinning of an X-point \edit1{(e.g. at $t=5\,\tau_A$)} or the presence of a nonlinear secondary tearing instability \edit1{(e.g. at $t=9.5\,\tau_A$)}. 
\begin{figure*}
 \includegraphics[width=0.335\textwidth,trim={0.0cm 0 0.55cm 0},clip]{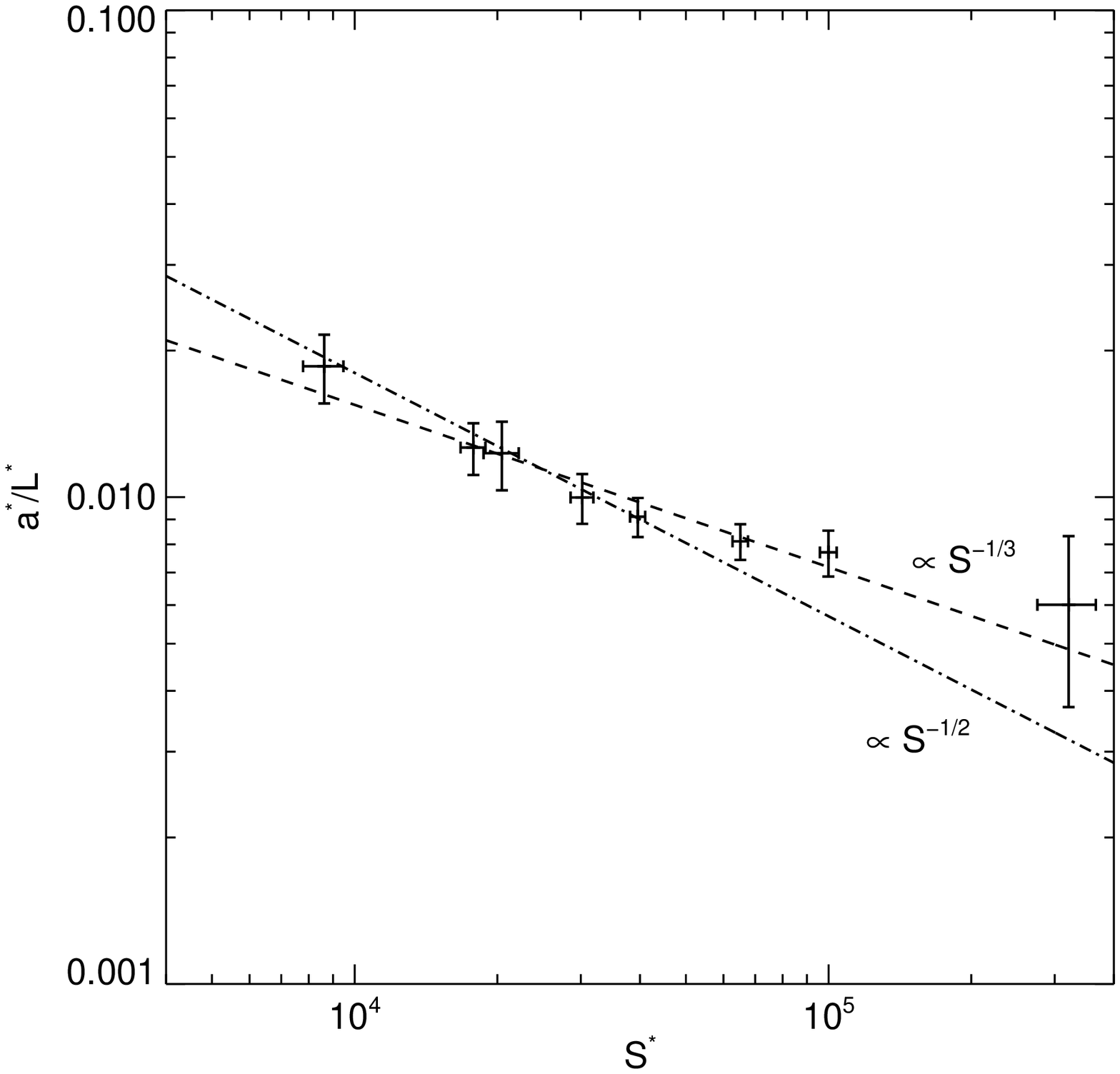}
 \includegraphics[width=0.335\textwidth,,trim={.0cm 0 0.55cm 0},clip]{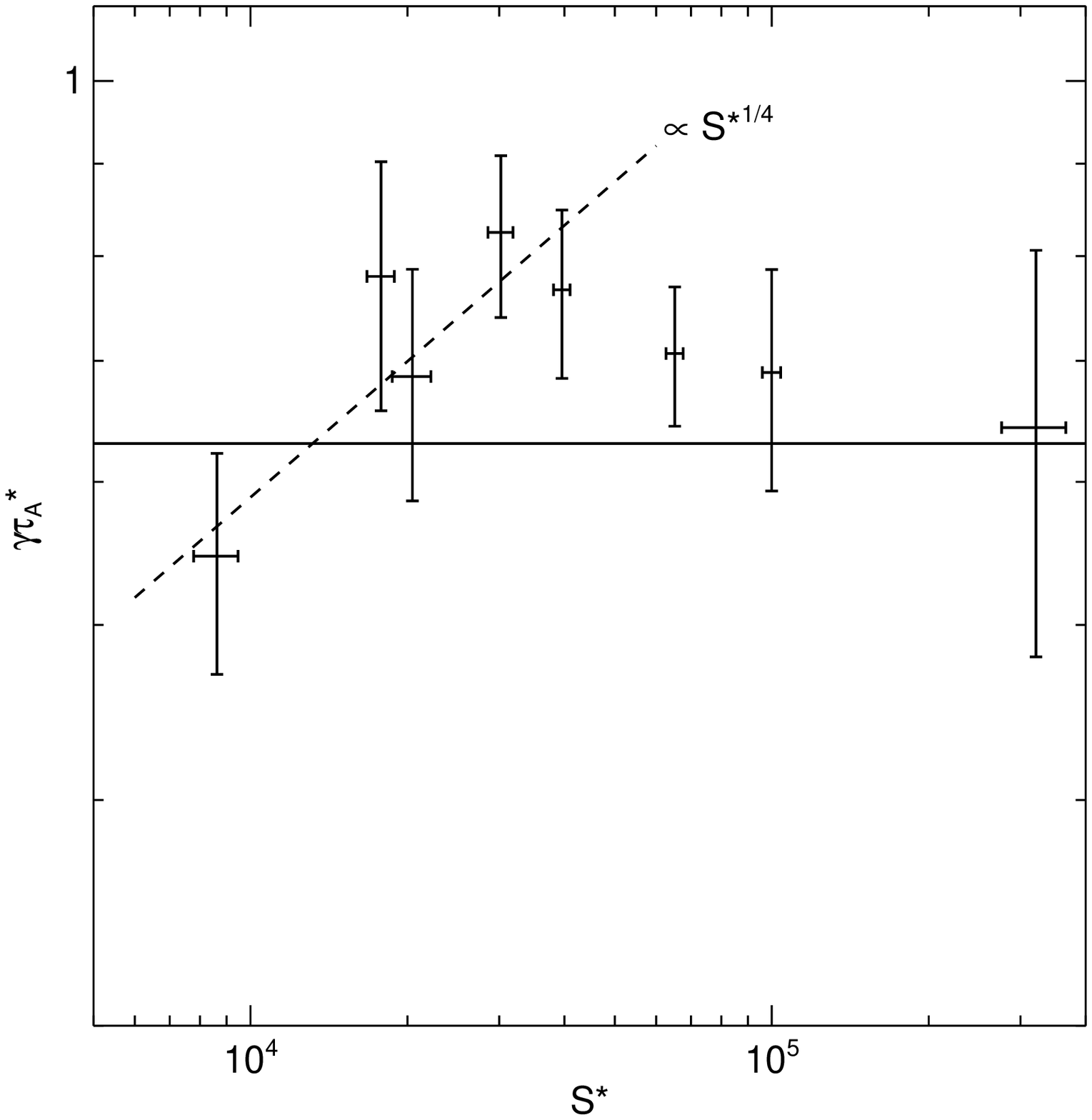}
 \includegraphics[width=0.335\textwidth,,trim={0.0cm 0 0.55cm 0},clip]{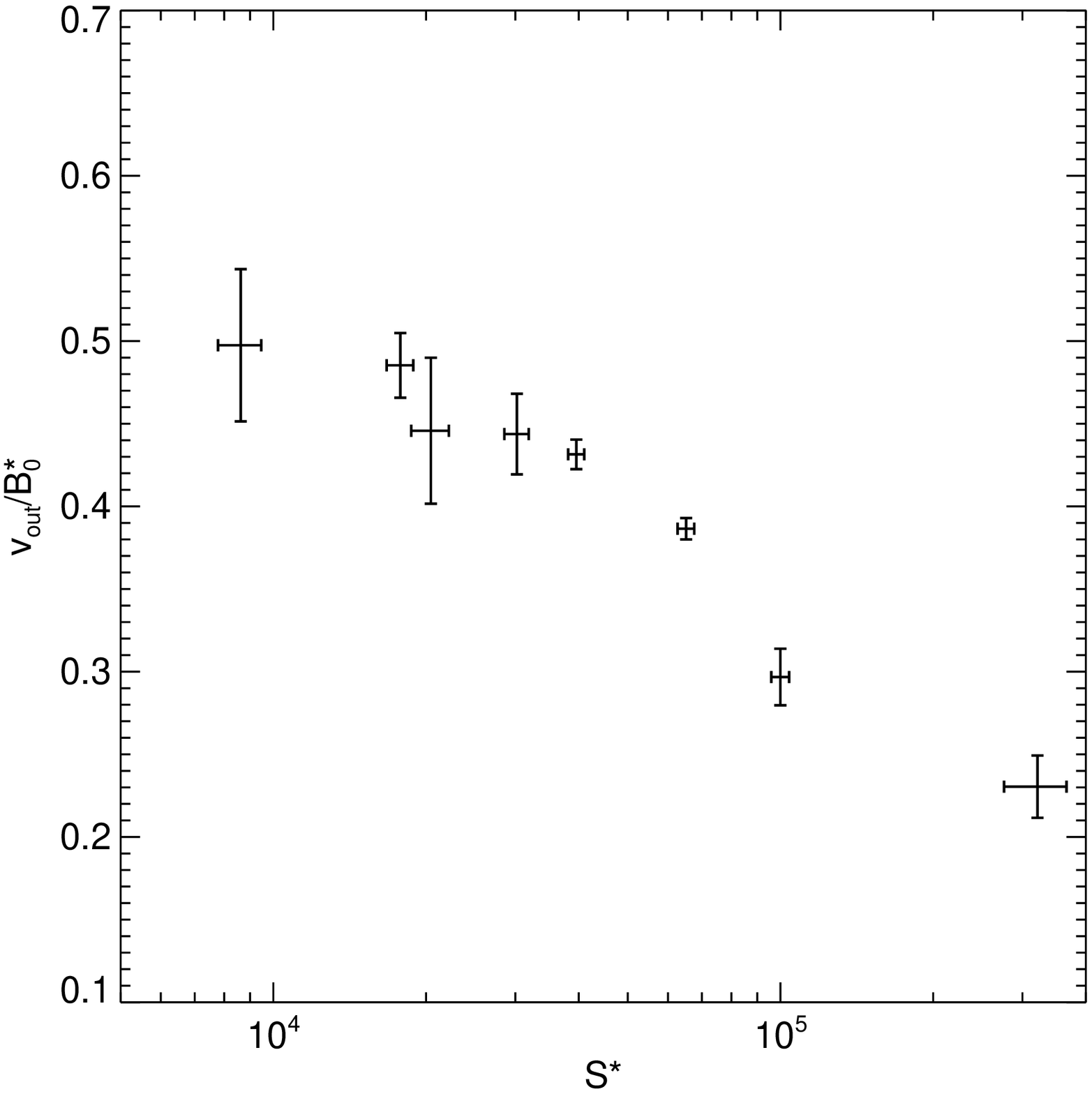}
 \caption{
Left panel: aspect ratio vs. local Lundquist number $S^*$ of secondary current sheets for a set of MHD simulations with \edit1{$S=10^5, \, 6.7\times 10^5, \, 8.0 \times 10^5, \, 10^6, \, 2\times 10^6, \, 5\times 10^6, \, 10^7, 10^8$ (Runs 5N to 8N and 12N to 15N respectively). Middle} panel: local reconnection rate $\gamma \tau_A^*$ vs. local Lundquist number $S^*$ for the secondary current sheets, normalized against the local Alfv\'en time $\tau_A^*$; the horizontal line denotes the theoretical linear growth rate of the most unstable mode for the ideal tearing instability ($0.63$, actually valid for $S^*\to\infty$). \edit1{ Right panel: amplitude $\rm{v}_\text{out}$ of the outflow velocity vs $S^*$ as measured at the edges of the secondary current sheets (i.e., at the coordinates $y_{cs} \pm L^*$), normalized with respect to the local Alfv\'en speed $c_A^*\simeq \rB_0^*$.}
}
\label{fig:scaling_secondary}
\end{figure*}

At the beginning of the simulation, at $t=0$, the fit correctly gives the thickness $a$ and the amplitude $\rB_0$ of the primary current sheet. As time proceeds, an X-point forms and then gets elongated due to the evacuation of nearby plasmoids. At about $t=7\,\tau_A$ a secondary current sheet has formed, since $\Delta a^* \ll a^*$ there. In the subsequent evolution, the current sheet further thins, but keeping a constant local Lundquist number, in this case $S_a^*\simeq 270$ (the dashed line in the figure). This happens because the magnetic field is kept frozen in the plasma inside the current sheet, since the diffusion time is much larger than the time scales of the thinning process. At $t=9~\tau_A$ a secondary linear tearing instability starts to develop inside the current sheet and the thinning stops concurrently.

The configuration of the secondary current sheet at this time clearly shows an almost perfect pressure equilibrium, as shown by the profiles of Fig.~\ref{fig:secondary_equilibrium}. Moreover, both an inflow perpendicular to the sheet and an outflow along its main direction are present, naturally formed because of the evacuation and merging of the plasmoids in the evolution of the primary reconnection process. As expected, the inflow is very weak \edit1{(but strong enough to counteract the diffusion of the magnetic field, since the plasma pile-up time $\tau_\text{up}=a^*/v_\text{in}^*$ associated with the inflow  is almost equal to the diffusion time $\tau_D = a^{*2}/\eta$)}, while the outflow peaks at roughly half of the local Alfv\'en speed. 
The secondary linear tearing instability that we have just described appears to be triggered by perturbations in the magnetic field with an amplitude of about $1\%$ with respect to $\rB_0^*$, hence it is bound to develop very rapidly and we actually witness the disruption of the secondary current sheet in less than an Alfv\'enic time. 

\edit1{To verify whether the dynamical formation and evolution of these secondary current sheets is insensitive of the initial equilibrium, we performed another run with the same parameters as run 8N but starting from a pressure equilibrium configuration. Results show that the dynamical evolution is the same, although the time of formation and triggering of the secondary tearing instability are shifted in time, due to a diffusion of the initial equilibrium field in the early times of the simulation \citep[][]{Landi:2015} that prolong the linear phase.}

The dynamics is qualitatively similar in all the MHD and Hall-MHD simulations we performed. In order to assess the scaling of the these secondary tearing instabilities with the local Lundquist number $S^*=L^*\rB_0^*/\eta$, here defined using the half length $L^*$ of the secondary current sheet as characteristic scale, we performed the same analysis on additional \edit1{seven MHD simulations (Run 5N to 7N and 12N to 15N of Table \ref{tab:sim_set}), by varying the global Lundquist number $S$ in the range $10^5 - 10^8$.} In Figure~\ref{fig:scaling_secondary} (left panel) we report, for each simulation, the aspect ratio $a^*/L^*$ of the secondary current sheet, calculated at the time when the thinning stops (different for each simulation), against $S^*$. The plot clearly shows that the scaling is consistent with that characteristic of the ideal tearing ($a^*/L^* \propto S^{* -1/3}$), although the SP scaling seems to be more appropriate for the lowest values of $S^*$. This may suggest the existence of two different regimes, in agreement with the findings of \citet{Huang:2017}. This scenario is also confirmed by looking at the reconnection rate $\gamma$ of the secondary current sheets. Figure~\ref{fig:scaling_secondary} (\edit1{middle} panel) shows that, once rescaled to the local Alfv\'en time $\tau_A^* = L^*/c_A^*$, the growth rate is compatible with the value $\gamma\simeq 0.63$ of the ideal tearing instability, with the exception of few points, that seem to be more compatible with the SP scaling
\edit2{(in spite of the large error bars)}. Note, however, that here the local Lunquist number is close to the threshold minimum value of $10^4$ requested to allow super-tearing modes \citep[see][for an exploration of lower values]{Shi:2018}. Moreover, the agreement with the critical scenario of the ideal tearing seems to be improving with increasing $S^*$, as expected, since we are moving toward the asymptotic regime ($S^*\rightarrow\infty$).

In this section we have demonstrated that, in the evolution of the nonlinear phase of MHD (and similarly for Hall-MHD) reconnection, secondary events occur inside the reconnecting sheet, which  \emph{spontaneously} adjust so to reach an ideal tearing regime: a local (inverse) aspect ratio of the secondary current sheet $a^*/L^* \sim S^{*-1/3}$ and a local growth rate of the linear tearing fastest mode $\gamma_\text{rec}\,\tau_A\simeq 0.63$, independent of the local Lundquist number when $S\to\infty$.

\section{Role of the Hall term}
\label{sec:second_hall}
\begin{figure}[t]
\includegraphics[width=\columnwidth]{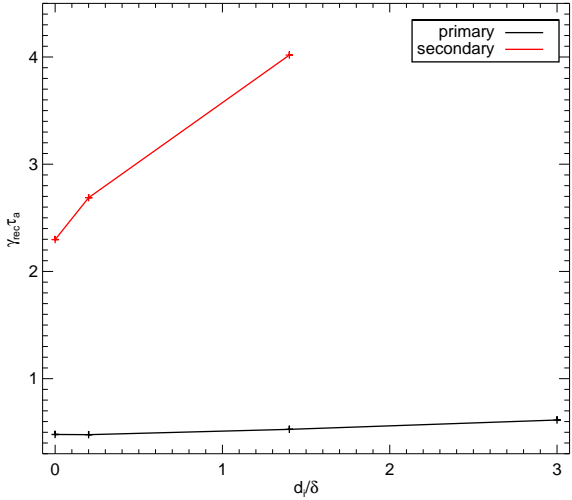}% rec3ndhall.eps}
\caption{Reconnection rates, calculated with an exponential fit of $B_{x,\mathrm{rms}}$, of primary (black) and secondary (red) current sheets, for simulations with  $S=10^6$ and $d_i/\delta=0.0, \, 0.2, \, 1.4, \, 3.0$ (Runs 8N-11N).}
\label{fig:recrates_hall}
\end{figure}

Even though for values of $d_i/\delta \lesssim 1$ the Hall term does not affect the reconnection rates of the primary current sheet, as we have already shown in Section \ref{sec:linear_phase}, the changes in the secondary reconnection events are substantial. Figure~\ref{fig:recrates_hall} shows a plot of $\gamma_\mathrm{rec}$ for the primary and the secondary reconnection events for simulations with  $S=10^6$ and different values of the ratio $d_i/\delta$ (Runs 8N-11N of Table \ref{tab:sim_set}). As we can see, the reconnection rate of the secondary events (red lines) increases by $17\%$ for a value $d_i/\delta=0.2$, and almost doubles for $d_i/\delta=1.4$. As a consequence, the evolution of the overall reconnection process is faster, the growth rate being up to five times the one of the MHD primary instability, and leads to the disruption of the current sheet in a correspondingly shorter time.
\edit1{The increasing with $\eta_H$ of the reconnection rate $\gamma^*\tau_A$ of these secondary events is due to the increase of the ratio  $d_i/\delta^*$, where $\delta^*$ is the inner resistive layer of the secondary current sheet. By rewriting Eq. (\ref{eq:di_delta}) for the quantities $a^*$ and $S_a^*$ retrieved by the fitting procedure (see previous section), we obtain a value of $d_i/\delta^* \simeq 1$ and $d_i/\delta^* \simeq 6$ for $d_i/\delta=0.2$ and $1.4$, respectively. }
\edit2{These results may be interpreted as the existence of a Hall super-tearing regime, in which the reconnection rate increases with the ratio $d_i/\delta^*$. Such interpretation is, however, incorrect.
Indeed,} \edit1{once renormalized to their local Alfv\'en time $\tau_A^*=L^*/\rB_0^*$, the reconnection rates of both cases are of the order of unity, due to the fact that the secondary sheet with the higher $d_i/\delta^*$ is shorther, and hence its local Alfv\'en time $\tau_A^*$ is smaller.} 

Moving to larger ratios $d_i/\delta$ (the case of $d_i/\delta=3$ is shown in the plot) we enter the strong Hall regime: the primary reconnection events become more and more violent, and the formation of secondary current sheets seems to be inhibited. The cause for the suppression of the secondary reconnection events in the strong Hall regime is unknown, however we can identify three possible explanations. 
The first possibility is that, for $d_i\gg\delta$, the linear growth rate of the secondary Hall tearing instability is so fast that it disrupts any forming current sheet before it can sufficiently grow to become dynamically important.
A second possibility is that the geometrical configuration (i.e., the quadrupolar structure) of a X-point in the Hall regime prevents the formation of a secondary current sheet.
Alternatively, the numerical resolution employed here may not be sufficient to reproduce the dynamics of the secondary reconnection events.

\section{Conclusions}

In this work we have presented a detailed study of the ideal tearing instability of thin current sheets in MHD and Hall-MHD plasmas, carried out by means of 2D compressible and fully nonlinear numerical simulations, along the same lines of \citet{Landi:2015} and \citet{DelZanna:2016}. Our results confirm that magnetic reconnection via the ideal tearing instability is indeed an efficient mechanism of energy conversion, which is as fast as the ideal Alfv\'en timescales in MHD, and even faster in the Hall regime.

In the MHD regime, after the ideal tearing instability saturated and the nonlinear phase has begun, we observed the onset of secondary reconnection events in newly formed current sheets, thinner by one order of magnitude than the initial current sheet. These secondary tearing instabilities are strongly super-Alfv\'enic, with reconnection rates $\gamma_\mathrm{rec}\tau_A\simeq 2.3$, \ie, five times faster than the main instability one. The net result is a much more violent reconnection process and a speed up in the disruption of the current sheet.  
Moreover, numerical simulations in the Hall-MHD regime, performed with increasing values of $\eta_H=d_i/L$, have shown that even though the Hall effect is negligible in the linear phase \edit1{for small values of $\eta_H$}, it considerably affects the secondary reconnection events in the nonlinear phase, by increasing the reconnection rates up to about $100\%$ (for $\eta_H=0.0014$, corresponding to $d_i/\delta = 1.4$ for our reference simulation with $S=10^6$) with respect to the pure MHD case and about ten times the reconnection rate of the linear phase. 
\edit1{This brings to a further speed up in the disruption of the whole current sheet.}
At higher values of $d_i/\delta$, the formation of secondary current sheets is not observed.

\edit1{Once renormalized to the local Alfv\'en time, the reconnection rate of the secondary events in the Hall regime becomes roughly constant and of the order of unity, which may suggest the existence of a modified Hall ideal tearing instability, as predicted by \citet{Pucci:2017}. However, the scaling of the measured aspect ratio $a^*/L^*$ is not consistent with their theoretical prediction $a^*/L^*\sim S^{*-1/3+0.29/2} (d_i/L^*)^{0.29}$, at least with the limited simulation dataset available in this study. A wider parameter study in the Hall regime, encompassing higher values of $S^*$, is required in order to assess the existence of this modified ideal tearing regime.}

Particular attention has been devoted to the study of the conditions under which the secondary instability takes place. 
Previous studies already identified and highlighted the properties of ideal tearing instabilities triggered in secondary current sheets, that had formed either in presence of an artificially induced collapse \citep{Tenerani:2015} or spontaneously \citep{Landi:2015,Landi:2017} from the primary current sheet.
Here we have quantitatively demonstrated for the first time that the new substructures, namely the thinning secondary current sheets formed among nearby X-points, \emph{spontaneously} tend to the critical aspect ratio proper of the ideal tearing, $a^*/L^*\sim S^{* -1/3}$ for high $S^*$, this time calculated on the \emph{local} spatial scales. 
In this phase the local Lundquist number remains constant and the sheet is in pressure balance with the external medium. Then the secondary instability fully develops, on timescales approaching the expected value $\gamma_\text{rec}\tau_A^*\simeq 0.63$, here using the shorter local value of $\tau_A^* \ll \tau_A^{\phantom{*}}$, thus on super-Alfv\'enic global timescales. This scenario has been investigated by performing several simulations varying the (global) Lundquist number in the range $S=10^5 - 10^8$: the ideal scaling for the locally formed secondary current sheets is retrieved for high $S^* (> 3\cdot10^4)$, and the match with the asymptotic value for the instability growth rate improves with increasing values of $S$, as expected. 
Instead, for moderate low $S^* (\lesssim 3\cdot10^4)$ a regime compatible with a SP scaling is observed.

\edit1{The existence of the two regimes \edit2{in our simulations} can be explained by the presence of the outflows which,  at moderate low $S^*$, are able to efficiently evacuate the tearing modes from the current sheet, thus stabilizing it \edit2{\citep[see, e.g.,][]{Ni:2010,Tenerani:2016,Shi:2018}.}
Below a critical threshold $S^*<S_c$ \edit2{\citep[that can be even smaller than $10^4$, ][]{Shi:2018}} the outflows effectively suppress the linear tearing instability and we retrieve a Sweet-Parker slow stationary reconnection . 
As the Lundquist number exceeds $S_c$, the current sheet becomes unstable. However, the outflow is still almost Alfv\'enic (see right panel of Fig. \ref{fig:scaling_secondary}) and provides some stabilization. 
At moderate low $S^*\simeq 10^4 $, the ideal tearing (hereafter IT) instability has a growth rate $\gamma_\mathrm{IT}\tau_A \lesssim 0.5$ that is too small to counteract the effect of the outflows. 
\edit2{Instead, a current sheet of SP aspect ratio has a growth rate $\gamma_\mathrm{SP}\tau_A \gg 1$ ($>4$ for $S^*=10^4$) much larger than the outflow evacuation rate. Moreover, it is easier for a thinning current sheet to further shrink to a SP ratio, since the thickness ($a_\mathrm{IT}$) of a IT sheet  is comparable to that  of a SP sheet ($a_\mathrm{SP}$) with the same $S^*$ ($a_\mathrm{IT}/a_\mathrm{SP}=S^{* 1/6}\simeq 4$ for $S^*=10^4$).}
As $S^*$ increases, the scale separation between the two configurations increases ($a_\mathrm{IT} \gg a_\mathrm{SP}$), and the outflow amplitude decreases (see right panel of Fig. \ref{fig:scaling_secondary}). Now a current sheet can be efficiently destroyed once it reached the IT aspect ratio, provided (as it is the case in our simulations) that the thinning process is slow compared to the local Alfv\'en time $\tau_A^*$, so that the current sheet is disrupted before it can thin further.
}

\edit2{The transition from SP scaling to IT has already been found by \citet{Huang:2017} (hereafter HU17), but with some differences. In their work, the change of regime} takes place at higher values of the Lundquist number (between $10^5$ and $10^6$). Moreover, they find a growth rate for the tearing instability which is much faster than 0.63, but still with the correct scalings in the two regimes. We believe that this discrepancy is only apparent, and it is due to the different normalization they used. In fact, in HU17 the current sheets form with a average half length $L^*=0.25$ (using our notation), which is correctly employed in the definition of $S^*=L^*c_A^*/\eta$. However, the Alfv\'en speed is set to the global one ($c_A^*=c_A^{\phantom{*}}=1$), without a strict check of its value,  unlike done in this work. This may lead to overestimating $S^*$. Moreover, the measured growth rates (see Table 1 of HU17) are normalized with respect to a global Alfv\'en time $\tau_A=1$,  and not with respect to the correct value $\tau_A^*=L^*/c_A^*$, that would be already four times smaller by using $L^*=0.25$.

Our detailed analysis on the reconnection dynamics of the secondary current sheets validates some predictions of \citet{Tenerani:2015}. For instance, we retrieved a thickness of the secondary current sheet that roughly corresponds to the width of the inner resistive layer of the primary current sheet $a^*\simeq \delta\sim S^{1/2}$. Moreover, in the MHD case with $S=10^6$, our measured value of the reconnection rate $\gamma^*\tau_A\simeq 2.3$  matches their theoretical estimate for the linear growth rate of the secondary tearing instability.

Outcomes of this work have potential applications for explaining the explosive events in the strongly magnetized space and astrophysical plasmas mentioned in the introduction, and further extend results of previous works of recursive magnetic reconnection \citep{Shibata:2001,Tenerani:2015,Singh:2019}. Moreover, the interplay between fast reconnection and turbulence can be crucial, as predicted by reconnection-mediated turbulence models \citep{Boldyrev:2012,Loureiro:2017,Mallet:2017} and by recent numerical simulations of the solar wind plasma retaining kinetic effects \citep{Franci:2017}, where the role of reconnection in driving the turbulent cascade at sub-ion scales through the destabilization of current sheets of thickness $a \simeq d_i$ is established. However, a recent study by \citet{Papini:2019} has demonstrated that the role of reconnection in shaping the spectrum of solar wind turbulence, including the change of slope at the ion inertial length $d_i$, may be captured even without resorting to (hybrid) particle-in-cell simulations, by just retaining the Hall effect within a macroscopic MHD description, as in the present study.

\section*{Acknowledgements}
The authors wish to acknowledge valuable exchanges of ideas at T.~Tullio. EP thanks Luca Franci and Daniele Del Sarto for fruitful discussion. SL and LDZ acknowledge support from the PRIN-MIUR project prot. 2015L5EE2Y {\it Multi-scale simulations of high-energy astrophysical plasmas}. This research was conducted with high performance computing (HPC) resources provided by the CINECA ISCRA initiative (grant HP10C2EARF and HP10B2DRR4).

\bibliography{biblio}
\bibliographystyle{aasjournal}

\end{document}